# A Hybrid Approach for COVID-19 Detection: Combining Wasserstein GAN with Transfer Learning

**Sumera Rounaq[1], Shahid Munir Shah[2], Mahmoud Aljawarneh[3]**

[1]Department of Computer Science, DHA Suffa University, Karachi, Pakistan

[2]Department of Computing, Hamdard University, Karachi, Pakistan

[3]Applied Science Private University, Amman, Jordan

**Abstract:**

Novel Coronavirus disease (COVID-19) has abruptly and undoubtedly changed the world as we know it at the end of the 2nd decade of the 21st century. COVID-19 is extremely contagious and its rapid growth has drawn attention towards its early diagnosis. Early diagnosis of COVID-19 enables health care professionals and government authorities to break the chain of transition and flatten the epidemic curve. With the number of cases accelerating across the developed world, COVID-19 induced Viral Pneumonia cases is a big challenge. Overlapping of COVID-19 cases with Viral Pneumonia and other lung infections with limited dataset and long training hours is a serious problem to cater. Limited amount of data often results in over-fitting models and due to this reason, model does not predict generalized results. To fill this gap, we proposed GAN based approach to synthesize images which later fed in to the deep learning models to classify images of COVID-19, Normal, and Viral Pneumonia. Specifically, customized Wasserstein GAN is proposed to generate 19% more Chest X-ray images as compare to the real images. This expanded dataset is then used to train four proposed deep learning models: VGG-16, ResNet-50, GoogLeNet and MNAST. The result showed that expanded dataset utilized deep learning models to deliver high classification accuracies. In particular, VGG-16 achieved highest accuracy of 99.17% among all four proposed schemes. Rest of the models like ResNet-50, GoogLeNet and MNAST delivered 93.9%, 94.49% and 97.75% testing accuracies respectively. Later, the efficiency of these models is compared with the state of art models on the basis of accuracy. Further, our proposed models can be applied to address the issue of scant datasets for any problem of image analysis.

## Section 1: Introduction

The coronavirus disease, commonly known as COVID-19, is a viral infection caused by the Severe Acute Respiratory Syndrome Coronavirus 2 (SARS-CoV-2). Emerging initially in Wuhan, China, this disease rapidly evolved into a global pandemic, impacting millions of individuals worldwide [1]. According to the World Health Organization (WHO), by the first half of 2022, the death toll had surpassed 6.2 million, with over 516 million confirmed cases globally [2].SARS-CoV-2 belongs to the same family as Severe Acute Respiratory Syndrome (SARS) and Middle East Respiratory Syndrome (MERS) [2].The most prevalent method for diagnosing COVID-19 is reverse transcription polymerase chain reaction (RT-PCR). Further, COVID-19 is classified as a respiratory illness, characterized by symptoms such as chest pain, high fever, dry cough, myalgia, and headache. The virus spreads through tiny droplets expelled from an infected person's mouth or nose [3]. The significant negative impacts of this disease underscore the necessity for effective measures to combat its spread. The rapid person-to-person transmission has prompted the implementation of stringent safety protocols, such as social distancing, to mitigate outbreaks [4]. As far as, remedial measures are concerned Computed Tomography (CT) and X-ray imaging have proven effective in examining lung diseases, including tuberculosis, pneumonia, infiltration, atelectasis, and COVID-19[5]. However, many underdeveloped and impoverished regions lack specialized human resources, impeding the use of these imaging technologies. Consequently, there has been a growing interest in computer-aided intelligent decision-making systems to automate the necessary processes.

Machine learning (ML) is increasingly utilized in medical imaging to enhance diagnostic accuracy, reduce analysis time and costs, and improve our understanding of diseases. A key advantage of ML in medical imaging is its capacity to automatically extract relevant features and patterns from large datasets, thus boosting diagnostic accuracy and efficiency. Deep learning (DL), a subset of ML, uses artificial neural networks with multiple layers to analyze complex data representations. This advanced approach has gained popularity in various fields, particularly in medical diagnosis, where DL methods significantly enhance image processing capabilities [6]. Detecting COVID-19 from chest X-rays is crucial for controlling the pandemic's spread, yet interpreting these images is challenging due to the virus's subtle and complex patterns. To address this issue, researchers employ optimization algorithms—mathematical methods designed to find the best solutions to problems. These algorithms can identify key features in the images, such as ground-glass opacities and consolidations, indicative of the virus. By optimizing these features, the algorithms improve the ability to detect COVID-19 in chest X-rays, facilitating earlier diagnosis and treatment [7].

To classify COVID-19 cases from the normal one or from pneumonia disease, many researchers relied on deep learning models [8, 9, and 10]. Despite their success in artificial intelligence applications [11], these models have not contributed to data expansion and require significant resources for training and validation due to the numerous parameters that need optimization. Further, ranging from medical health images classification to object detection, Convolutional Neural Network is being chosen, since it is comprised of assemblies of nodes known as neurons. Mostly models are comprised of two types: Pre-trained Models and Non-Pre-trained Models. Non-Pre-trained models are prone to over fitting and massive amount of data is required to train these models. In contrast, Pre-trained models resolved the problem of overfitting since they get trained using publicly available datasets such as Image Net. [14]. Keeping in view all of the above mentioned challenges our contributions are as follows:

1- To overcome the issue of overfitting and enhance model generalization, customized data augmentation technique is proposed i.e. Wasserstein GAN (WGAN) using gradient penalty.
2- WGAN architecture also contributed in alleviation of mode collapse and vanishing gradient.
3- To achieve better performance in the detection of COVID-19 cases, pre-trained models (VGG-16, ResNet-50, GoogLeNet and MNAST) are employed.
4- To extract high quality features and capture essential patterns in Chest X-Ray (CXR) images, pre-trained models are suggested to train using feature extraction techniques.
5- Fusion of Wasserstein GAN and pre-trained models are compared with the state-of-the-art models to achieve optimal solution.

The remainder of the paper is structured as follows: Section 2 outlines the latest relevant research related to the study. Further, Section 3 explains the methodology of the proposed approach, including architecture and parametric configuration of the model. Section 4 presents the results and their discussion, analyzing the outcomes of the experiments and their implications compared to existing methods. Section 5 discusses the limitations of the proposed approach, addressing any constraints or challenges encountered and suggesting potential areas for further research to address these limitations and expand upon the findings.

**Section 2: Literature Review**

Detection of COVID-19 is immensely important in order to secure people's lives. In this regard, significant approaches have been taken by scientists which have delivered promising results. To begin with, Bhattacharyya, Abhijit, et.al. (2022) [15] proposed deep learning techniques by dividing the dataset into three categories as Normal, COVID-19, and Pneumonia. To classify the images, they relied on five CNN variants namely, VGG-16, VGG-19, CNN, DenseNet-169, and Dense Net 201. Moreover, for feature detection and description of images they deployed SIFT (Scale-Invariant Feature Transform) and BRISK (Binary Robust Invariant Scalable Key Points). This step assigns a consistent orientation to each key point based on local image gradients. In addition to it, these two techniques operate on multiple scales to detect key points that are invariant to scale changes. For image segmentation purposes, Fedoruk, Oleksandr, et al. (2024) [16] employed the K-Means clustering algorithm and divided the images into 128 clusters. StyleGAN2-ADA architecture was suggested to augment the dataset which was classified into four categories i.e., COVID-19, Lung Opacity, Viral Pneumonia, and healthy lungs. Following the augmentation they developed two classification network architectures namely Inception-V3 and Efficient Net-B0. To delineate the intensity variations between a chest X-ray (CXR) and its rib-suppressed counterpart Han, Luyi, et al. (2022) [17] employed a residual map and to forecast this residual map within the CXR framework, they employed a disentanglement technique, segregating the image into distinct features that capture both structural and contrast-specific elements. Their proposed Rib Suppression GAN (RSGAN) outperformed existing rib suppression methods in terms of image quality, and by combining CXR with rib-suppressed output yields improved performance in lung disease classification and tuberculosis area detection.

In study [18], Togacar et al. (2021) utilized the LIME (Local Interpretable Model-agnostic explanations) approach for the identification of COVID-19. They employed CT radiological images from two distinct groups. The first group, termed Cov-Pne-Bac (CPB), was utilized for classifying COVID-19, H1N1, and Bacterial pneumonia cases. The researchers employed three residual networks (ResNet18, ResNet50, and ResNet101) to classify the parameters. For dataset preprocessing, Grade-CAM and Fourier Transform techniques were employed. Despite the reliance of supervised models on extensive labeled data, which can be challenging and costly to obtain, particularly for emerging diseases, Yadav, Pooja, et al. (2021) [19] present a novel approach in this research. They proposed a deep unsupervised framework for lung disease classification using chest CT and X-ray images. Their framework, Lung-GANs, comprises multiple-layer generative adversarial networks designed to learn interpretable representations of lung disease images solely from unlabeled data. Leveraging the learned lung features, they trained a Support Vector Machine (SVM) and a Stacking Classifier. Experimental results demonstrated in their research work surpassed the current state-of-the-art unsupervised models in lung disease classification. Across six large-scale publicly available lung disease datasets used in study, their proposed model achieved an accuracy ranging from 94% to 99.5%.

Vashisht, Sanchit, et al. (2023) [20] considered three categories: normal, viral, and bacterial pneumonia, utilizing chest X-rays, the primary diagnostic tool for these ailments, in the form of scanned images. By aggregating images from online repositories and employing the GAN algorithm, the dataset size was augmented, facilitating more robust model training and testing. Utilizing a combination of CNN and GAN approaches, disease classification achieved an accuracy of 98.76%, surpassing other techniques. Consequently, this proposed system enhanced pneumonia detection efficiency, even with limited datasets. In their study, Rahimzadeh et al. (2020) [21] conducted COVID-19 classification utilizing a concatenated approach of CNN with a foundation based on Xception and ResNet50V2 models, leveraging chest X-ray images. Their proposed methodology involved eight training phases, wherein the dataset was categorized into three classes: COVID-19, Pneumonia, and Normal. The proposed system achieved an impressive accuracy of 99.56% and a recall of 80.53% specifically for COVID-19 cases.

Vinayakumar Ravi et al. (2022) [22] introduced a novel large-scale learning approach that employed a stacked ensemble meta-classifier and deep learning-based feature fusion for COVID-19 classification. Initially, features were extracted from the penultimate layer (global average pooling) of pre-trained EfficientNet-based models, followed by dimensionality reduction using kernel principal component analysis (PCA). Subsequently, a feature fusion technique merged the extracted features from various sources. Finally, a two-stage classification process was employed. In the first stage, random forest and support vector machine (SVM) algorithms were utilized for prediction, and their outputs were aggregated for input into the second stage. The second stage involved a logistic regression classifier, which categorized CT and CXR data samples into either COVID-19 or Non-COVID-19. The proposed model underwent testing using large publicly available CT and CXR datasets, and its performance was compared against various existing CNN-based pretrained models. An efficient deep convolutional generative adversarial network and convolutional neural network (DGCNN) was designed by Laddha et al. (2022) [23] for diagnosing COVID-19 suspected subjects. The deep convolutional generative adversarial network (DGAN) utilized two networks trained adversarially, with one generating fake images and the other discerning between them. In parallel, a convolutional neural network (CNN) was leveraged for classification tasks. The dataset used in this study is well-structured, comprising four distinct classes: COVID-19, normal, pneumonia viral, and pneumonia bacterial. In total, the dataset consists of 306 images, which are further divided into 270 training images and 36 test images. Their proposed model outperformed state-of-the-art models and achieved 98.89% accuracy.

The Abdusalomov et al. (2023) [24] utilized feature maps of images extracted through various methods or distribution of image sets. Subsequently, the proximity of synthetic images to the real set was assessed using diverse distance metrics. A novel method was proposed to quantitatively and qualitatively evaluate synthetic images, combining two approaches: FMD and CNN-based evaluation methods. These estimation methods were compared against the FID method, revealing that while the FMD method excels in speed, the CNN method demonstrates superior accuracy in estimation. To assess the reliability of these methods, a dataset comprising different real images was examined. In their study, Kora Venu et al. (2020) [25] cyclegan, pix2pix, and deep pix2pix generative adversarial networks (GAN) were employed for generating synthetic medical images. To enhance the pix2pix model's performance in synthesizing images from one view to another, the authors developed a generator architecture inspired by the U-Net design. Three models, including cyclegan, pix2pix, and the proposed deep pix2pix, were compared for their ability to generate synthetic images for qualitative analysis. Results revealed that the proposed deep pix2pix model exhibited the most accurate results among the models used in the experiment, while the cyclegan model showed the poorest qualitative analysis performance.Gulakala et al. [26] developed two novel CNN architectures to classify COVID-19, healthy, and pneumonia-affected chest X-rays. To enhance the dataset, they employed a Progressively Growing Generative Adversarial Network (PGGAN) to generate synthetic and augmented data. Further to ensure the robustness of their model, Sohaib Asif et al. (2024) [27] suggested an ensemble technique by employing fourteen architectures including VGG, DenseNet, InceptionResNetV2, ResNetV2 (50, 101, 152), InceptionV3, NasNetMobile, Xception and MobileNet. To gain a deeper understanding of features and highlight the specific lung regions for accurate detection, an innovative approach of GradCAM was utilized. Their ensemble model achieved 99.03% accuracy for 3-class classification and 99.02% accuracy for 4-class classification.

**Section 3: Methodology**

The proposed method is composed of two phases, first phase is based on GAN which is further divided in to two neural networks i.e., Generator and Discriminator and second phase is based on transfer learning techniques. To elaborate more, GAN technique is used in preprocessing phase to expand dataset which then further fed in to transfer learning techniques to classify Chest X-ray images. Overall structure of the methodology is depicted in the block diagram Fig. [1]. As shown in the figure, original dataset [28] is comprised of Chest X-ray images belongs to three classes i.e., COVID-19, Normal and Viral Pneumonia. To ensure robustness of algorithm, this research study included Viral Pneumonia images along with COVID-19 images. Since chest X-ray images of COVID-19 cases usually have overlapping symptoms with Viral Pneumonia cases [29] so designed scheme of classes would help classifier to classify accurately.

**Fig. [3.1] Block Diagram of proposed methodology**

### 3.1 : Data Acquisition

Chest X-ray dataset [28] is used in this study which is created by post-doctoral fellow of University of Montreal, Dr. Joseph Cohen. Dataset [28] is available on GitHub repository and all the AP (anteroposterior) chest radiographs are extracted using metadata csv file available in repository. All Chest X-ray images are classified into COVID-19, Normal and Viral Pneumonia classes and are converted into PNG (Portable Graphics Format). Table [3.1] represents number of real images extracted for each class from dataset [28]. Accordingly, separate images have been taken for training and testing purpose. Moreover, total of 438 images are used to generate images for training dataset and, on the other hand, 150 images are used to generate images for test dataset. Samples of real images used in this study are shown in Fig [3.2].

| Dataset/Classes | COVID -19 | Normal | Viral Pneumonia | Total |
|---|---|---|---|---|
| **Train** | 146 | 146 | 146 | **438** |
| **Test** | 50 | 50 | 50 | **150** |
| **Total** | **196** | **196** | **196** | **588** |

Table [3.1] Number of real images for each class from dataset [28]

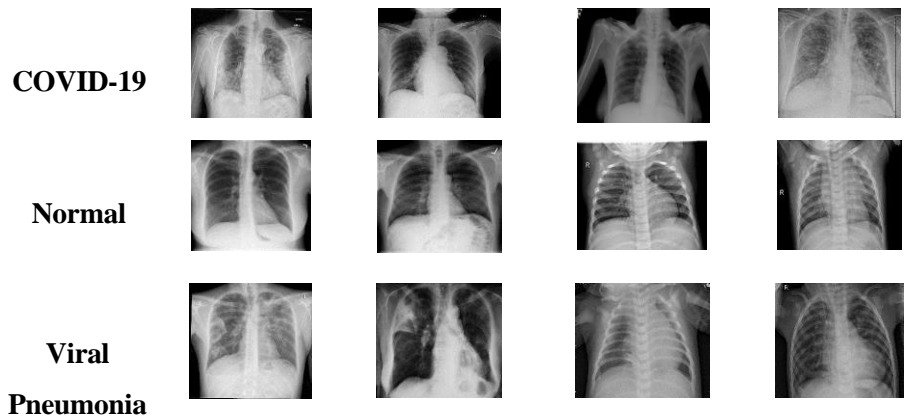

Fig [3.2] Samples of real images from dataset [28]

### 3.2 : Proposed WGAN-GP Architecture

Architecture shown in Fig [3.3] is comprised of generator and critic where generator is responsible to generate new plausible images from training dataset and critic's job is to classify those images as fake or real. Generator takes noise vector (z) from latent space as input and generate images. Deep convolutional neural network is implied as baseline for GANs. Traditional convolutional neural networks use pooling layers to reduce the size of an image while GANs requires an inverse operation to generate more detailed output from salient features. For this purpose, GANs employs up sampling. Up-sampling has the opposite effect of pooling. Given the lower resolution image, up-sampling has a goal of outputting one that has higher resolution. It actually requires inferring values for the additional pixels and like pooling layers, up sampling layers don't have any learnable parameters. Along with up sampling layers, GANs also use transpose convolution layers to enlarge the size of the generated output.

In our proposed WGAN architecture, initial image is given of 128 * 1 * 1 resolution where 128 is the dimension of the noise vector and final output of 1 * 128 * 128 is generated by the generator. This generated image is then given as input to critic along with real images and in critic down sampling is performed and image is classified as real or fake. Intermediate image is created by computing gradients. Computation of gradients is done by calculating the fake and real image values using epsilon and add them together. This mixed image is then used to calculate critic's output. Finally, the gradient of the critic score is computed on the mixed images (output) with respect to the pixels of the mixed images (input). For this entire process critic is updated 5 times per each time generator is updated.

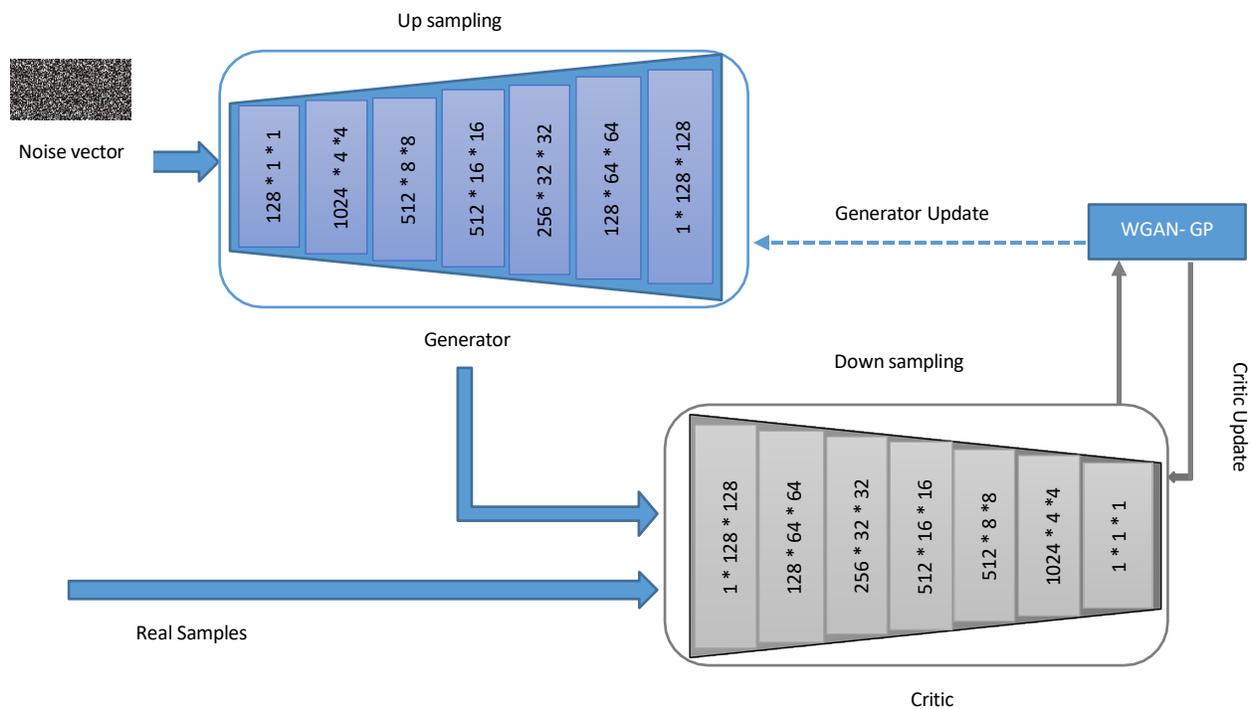

**Fig. [3.3] An overview of Wasserstein GAN (WGAN) architecture**

## 3.3 : Proposed Transfer Learning Architectures

In this research study, unlike traditional model training approach, where models are trained for each task independently, we applied four transfer learning approaches to classify the images as COVID-19, Normal and Viral Pneumonia.

### 3.3.1 VGG-16:

VGG -16 [30] focuses on just having convolutional layers that are using 3 x 3 filters with a stride of one followed by max pooling layers of size 2 x 2. It is a very simple neural network architecture with extensive capabilities to produce high accuracies as reported by research study [16]. This architecture starts with 224 x 224 x 3 dimensions. First two layers use 64 filters and [Conv 64 * 2] in diagram [3.4] shows 64 filters are applying on two convolutional layers. Then by using pooling on the image height and width of an image reduces to half and number of filters increases to double. In this architecture, convolution operation starts with 64 filters then doubles to 128 to 256 and to 512. This doubling through every stack of convolutional layers was another simple principle used to design the architecture of this network.

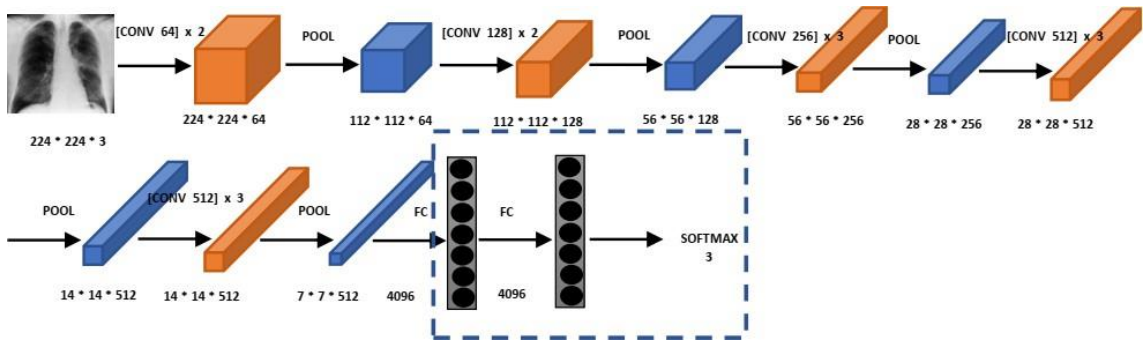

**Fig. [3.4] Architecture of VGG -16**

It is a deeper network in which trained parameters counts to approx. 138 million. In addition to it, number 16 in VGG – 16 refers to number of layers which have trainable parameters. As shown in Fig. [3.4] 13 convolutional layers and 3 layers at the end of the network makes the total count of 16.

**Architecture of the VGG-16 model:**

```
----------------------------------------------------------------
        Layer (type)               Output Shape         Param #
================================================================
            Conv2d-1         [-1, 64, 224, 224]           1,792
       BatchNorm2d-2         [-1, 64, 224, 224]             128
              ReLU-3         [-1, 64, 224, 224]               0
            Conv2d-4         [-1, 64, 224, 224]          36,928
       BatchNorm2d-5         [-1, 64, 224, 224]             128
              ReLU-6         [-1, 64, 224, 224]               0
         MaxPool2d-7         [-1, 64, 112, 112]               0
            Conv2d-8        [-1, 128, 112, 112]          73,856
       BatchNorm2d-9        [-1, 128, 112, 112]             256
             ReLU-10        [-1, 128, 112, 112]               0
           Conv2d-11        [-1, 128, 112, 112]         147,584
      BatchNorm2d-12        [-1, 128, 112, 112]             256
             ReLU-13        [-1, 128, 112, 112]               0
        MaxPool2d-14          [-1, 128, 56, 56]               0
           Conv2d-15          [-1, 256, 56, 56]         295,168
      BatchNorm2d-16          [-1, 256, 56, 56]             512
             ReLU-17          [-1, 256, 56, 56]               0
           Conv2d-18          [-1, 256, 56, 56]         590,080
      BatchNorm2d-19          [-1, 256, 56, 56]             512
             ReLU-20          [-1, 256, 56, 56]               0
           Conv2d-21          [-1, 256, 56, 56]         590,080
      BatchNorm2d-22          [-1, 256, 56, 56]             512
             ReLU-23          [-1, 256, 56, 56]               0
        MaxPool2d-24          [-1, 256, 28, 28]               0
           Conv2d-25          [-1, 512, 28, 28]       1,180,160
      BatchNorm2d-26          [-1, 512, 28, 28]           1,024
             ReLU-27          [-1, 512, 28, 28]               0
           Conv2d-28          [-1, 512, 28, 28]       2,359,808
      BatchNorm2d-29          [-1, 512, 28, 28]           1,024
             ReLU-30          [-1, 512, 28, 28]               0
           Conv2d-31          [-1, 512, 28, 28]       2,359,808
      BatchNorm2d-32          [-1, 512, 28, 28]           1,024
             ReLU-33          [-1, 512, 28, 28]               0
        MaxPool2d-34          [-1, 512, 14, 14]               0
           Conv2d-35          [-1, 512, 14, 14]       2,359,808
      BatchNorm2d-36          [-1, 512, 14, 14]           1,024
             ReLU-37          [-1, 512, 14, 14]               0
           Conv2d-38          [-1, 512, 14, 14]       2,359,808
      BatchNorm2d-39          [-1, 512, 14, 14]           1,024
             ReLU-40          [-1, 512, 14, 14]               0
           Conv2d-41          [-1, 512, 14, 14]       2,359,808
      BatchNorm2d-42          [-1, 512, 14, 14]           1,024
             ReLU-43          [-1, 512, 14, 14]               0
        MaxPool2d-44            [-1, 512, 7, 7]               0
AdaptiveAvgPool2d-45            [-1, 512, 7, 7]               0
           Linear-46                 [-1, 4096]     102,764,544
             ReLU-47                 [-1, 4096]               0
          Dropout-48                 [-1, 4096]               0
           Linear-49                 [-1, 4096]      16,781,312
             ReLU-50                 [-1, 4096]               0
```

```
         Dropout-51                    [-1, 4096]                      0
          Linear-52                       [-1, 3]                 12,291
            VGG-53                        [-1, 3]                      0
================================================================
Total params: 134,281,283
Trainable params: 12,291
Non-trainable params: 134,268,992
----------------------------------------------------------------
Input size (MB): 0.57
Forward/backward pass size (MB): 322.13
Params size (MB): 512.24
Estimated Total Size (MB): 834.95
```

**Table [3.2] Proposed Architecture of the VGG-16 model**

As shown in table [3.2] the model begins with an input layer that accepts an image of size 3×224×224. It features multiple convolutional layers organized by output dimensions: Layer Group 1 includes two convolutional layers with 64 filters each, followed by a max pooling layer; Layer Group 2 includes two convolutional layers with 128 filters each, followed by a max pooling layer; Layer Group 3 consists of three convolutional layers with 256 filters each, followed by a max pooling layer; Layer Group 4 has three convolutional layers with 512 filters each, followed by a max pooling layer; Layer Group 5 also includes three convolutional layers with 512 filters each, followed by a max pooling layer. Each convolutional layer is succeeded by a Batch Normalization layer and a ReLU activation function. Max pooling layers reduce the spatial dimensions of the feature maps after each group of convolutional layers. Following these, the model includes fully connected layers: the first fully connected layer reduces features to 4096 dimensions, followed by a ReLU activation and a Dropout layer; the second fully connected layer also reduces features to 4096 dimensions, followed by a ReLU activation and a Dropout layer; and the final fully connected layer reduces the output to 3 dimensions, corresponding to the classification of 3 classes. The final layer outputs the class scores for these 3 classes. During training, only the parameters of the final classification layer (12,291 parameters) are trainable, while the parameters of the pre-trained VGG-16 backbone (134,268,992 parameters) are frozen. The total estimated memory usage of the model is approximately 834.95 MB.

### 3.3.2 ResNet-50:

Saturation and degradation of accuracy of training models is a very common problem which results due to vanishing and exploding gradients. To address this issue, study [31] introduced the concept of residual learning shown in Fig. [3.5]. Residual networks learn about skip connections which allows us to take the activation from one layer and feed it to another layer even in much deeper neural networks. ResNet50 is a variant of residual networks which comprises of 48 convolutional layers, 1 maximum pooling and 1 average pooling layer.

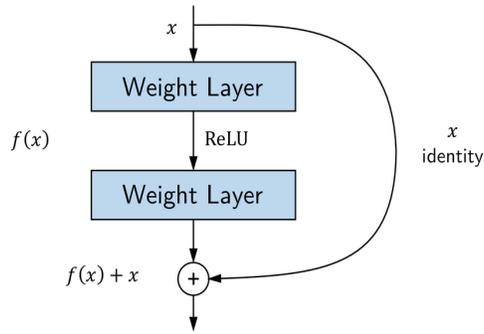

**Fig [3.5] Architecture of Residual Network.**

**Architecture of the ResNet 50 model:**

```
----------------------------------------------------------------
        Layer (type)               Output Shape         Param #
================================================================
            Conv2d-1         [-1, 64, 112, 112]           9,408
       BatchNorm2d-2         [-1, 64, 112, 112]             128
              ReLU-3         [-1, 64, 112, 112]               0
         MaxPool2d-4           [-1, 64, 56, 56]               0
            Conv2d-5           [-1, 64, 56, 56]           4,096
       BatchNorm2d-6           [-1, 64, 56, 56]             128
              ReLU-7           [-1, 64, 56, 56]               0
            Conv2d-8           [-1, 64, 56, 56]          36,864
       BatchNorm2d-9           [-1, 64, 56, 56]             128
             ReLU-10           [-1, 64, 56, 56]               0
           Conv2d-11          [-1, 256, 56, 56]          16,384
      BatchNorm2d-12          [-1, 256, 56, 56]             512
           Conv2d-13          [-1, 256, 56, 56]          16,384
      BatchNorm2d-14          [-1, 256, 56, 56]             512
             ReLU-15          [-1, 256, 56, 56]               0
       Bottleneck-16          [-1, 256, 56, 56]               0
           Conv2d-17           [-1, 64, 56, 56]          16,384
      BatchNorm2d-18           [-1, 64, 56, 56]             128
             ReLU-19           [-1, 64, 56, 56]               0
           Conv2d-20           [-1, 64, 56, 56]          36,864
      BatchNorm2d-21           [-1, 64, 56, 56]             128
             ReLU-22           [-1, 64, 56, 56]               0
           Conv2d-23          [-1, 256, 56, 56]          16,384
      BatchNorm2d-24          [-1, 256, 56, 56]             512
             ReLU-25          [-1, 256, 56, 56]               0
       Bottleneck-26          [-1, 256, 56, 56]               0
           Conv2d-27           [-1, 64, 56, 56]          16,384
      BatchNorm2d-28           [-1, 64, 56, 56]             128
             ReLU-29           [-1, 64, 56, 56]               0
           Conv2d-30           [-1, 64, 56, 56]          36,864
      BatchNorm2d-31           [-1, 64, 56, 56]             128
             ReLU-32           [-1, 64, 56, 56]               0
           Conv2d-33          [-1, 256, 56, 56]          16,384
      BatchNorm2d-34          [-1, 256, 56, 56]             512
             ReLU-35          [-1, 256, 56, 56]               0
       Bottleneck-36          [-1, 256, 56, 56]               0
```

```
        Conv2d-37         [-1, 128, 56, 56]          32,768
   BatchNorm2d-38         [-1, 128, 56, 56]             256
          ReLU-39         [-1, 128, 56, 56]               0
        Conv2d-40         [-1, 128, 28, 28]         147,456
   BatchNorm2d-41         [-1, 128, 28, 28]             256
          ReLU-42         [-1, 128, 28, 28]               0
        Conv2d-43         [-1, 512, 28, 28]          65,536
   BatchNorm2d-44         [-1, 512, 28, 28]           1,024
        Conv2d-45         [-1, 512, 28, 28]         131,072
   BatchNorm2d-46         [-1, 512, 28, 28]           1,024
          ReLU-47         [-1, 512, 28, 28]               0
    Bottleneck-48         [-1, 512, 28, 28]               0
        Conv2d-49         [-1, 128, 28, 28]          65,536
   BatchNorm2d-50         [-1, 128, 28, 28]             256
          ReLU-51         [-1, 128, 28, 28]               0
        Conv2d-52         [-1, 128, 28, 28]         147,456
   BatchNorm2d-53         [-1, 128, 28, 28]             256
          ReLU-54         [-1, 128, 28, 28]               0
        Conv2d-55         [-1, 512, 28, 28]          65,536
   BatchNorm2d-56         [-1, 512, 28, 28]           1,024
          ReLU-57         [-1, 512, 28, 28]               0
    Bottleneck-58         [-1, 512, 28, 28]               0
        Conv2d-59         [-1, 128, 28, 28]          65,536
   BatchNorm2d-60         [-1, 128, 28, 28]             256
          ReLU-61         [-1, 128, 28, 28]               0
        Conv2d-62         [-1, 128, 28, 28]         147,456
   BatchNorm2d-63         [-1, 128, 28, 28]             256
          ReLU-64         [-1, 128, 28, 28]               0
        Conv2d-65         [-1, 512, 28, 28]          65,536
   BatchNorm2d-66         [-1, 512, 28, 28]           1,024
          ReLU-67         [-1, 512, 28, 28]               0
    Bottleneck-68         [-1, 512, 28, 28]               0
        Conv2d-69         [-1, 128, 28, 28]          65,536
   BatchNorm2d-70         [-1, 128, 28, 28]             256
          ReLU-71         [-1, 128, 28, 28]               0
        Conv2d-72         [-1, 128, 28, 28]         147,456
   BatchNorm2d-73         [-1, 128, 28, 28]             256
          ReLU-74         [-1, 128, 28, 28]               0
        Conv2d-75         [-1, 512, 28, 28]          65,536
   BatchNorm2d-76         [-1, 512, 28, 28]           1,024
          ReLU-77         [-1, 512, 28, 28]               0
    Bottleneck-78         [-1, 512, 28, 28]               0
        Conv2d-79         [-1, 256, 28, 28]         131,072
   BatchNorm2d-80         [-1, 256, 28, 28]             512
          ReLU-81         [-1, 256, 28, 28]               0
        Conv2d-82         [-1, 256, 14, 14]         589,824
   BatchNorm2d-83         [-1, 256, 14, 14]             512
          ReLU-84         [-1, 256, 14, 14]               0
        Conv2d-85        [-1, 1024, 14, 14]         262,144
   BatchNorm2d-86        [-1, 1024, 14, 14]           2,048
        Conv2d-87        [-1, 1024, 14, 14]         524,288
   BatchNorm2d-88        [-1, 1024, 14, 14]           2,048
          ReLU-89        [-1, 1024, 14, 14]               0
    Bottleneck-90        [-1, 1024, 14, 14]               0
```

```
         Conv2d-91         [-1, 256, 14, 14]         262,144
    BatchNorm2d-92         [-1, 256, 14, 14]             512
           ReLU-93         [-1, 256, 14, 14]               0
         Conv2d-94         [-1, 256, 14, 14]         589,824
    BatchNorm2d-95         [-1, 256, 14, 14]             512
           ReLU-96         [-1, 256, 14, 14]               0
         Conv2d-97        [-1, 1024, 14, 14]         262,144
    BatchNorm2d-98        [-1, 1024, 14, 14]           2,048
           ReLU-99        [-1, 1024, 14, 14]               0
    Bottleneck-100        [-1, 1024, 14, 14]               0
        Conv2d-101         [-1, 256, 14, 14]         262,144
   BatchNorm2d-102         [-1, 256, 14, 14]             512
          ReLU-103         [-1, 256, 14, 14]               0
        Conv2d-104         [-1, 256, 14, 14]         589,824
   BatchNorm2d-105         [-1, 256, 14, 14]             512
          ReLU-106         [-1, 256, 14, 14]               0
        Conv2d-107        [-1, 1024, 14, 14]         262,144
   BatchNorm2d-108        [-1, 1024, 14, 14]           2,048
          ReLU-109        [-1, 1024, 14, 14]               0
    Bottleneck-110        [-1, 1024, 14, 14]               0
        Conv2d-111         [-1, 256, 14, 14]         262,144
   BatchNorm2d-112         [-1, 256, 14, 14]             512
          ReLU-113         [-1, 256, 14, 14]               0
        Conv2d-114         [-1, 256, 14, 14]         589,824
   BatchNorm2d-115         [-1, 256, 14, 14]             512
          ReLU-116         [-1, 256, 14, 14]               0
        Conv2d-117        [-1, 1024, 14, 14]         262,144
   BatchNorm2d-118        [-1, 1024, 14, 14]           2,048
          ReLU-119        [-1, 1024, 14, 14]               0
    Bottleneck-120        [-1, 1024, 14, 14]               0
        Conv2d-121         [-1, 256, 14, 14]         262,144
   BatchNorm2d-122         [-1, 256, 14, 14]             512
          ReLU-123         [-1, 256, 14, 14]               0
        Conv2d-124         [-1, 256, 14, 14]         589,824
   BatchNorm2d-125         [-1, 256, 14, 14]             512
          ReLU-126         [-1, 256, 14, 14]               0
        Conv2d-127        [-1, 1024, 14, 14]         262,144
   BatchNorm2d-128        [-1, 1024, 14, 14]           2,048
          ReLU-129        [-1, 1024, 14, 14]               0
    Bottleneck-130        [-1, 1024, 14, 14]               0
        Conv2d-131         [-1, 256, 14, 14]         262,144
   BatchNorm2d-132         [-1, 256, 14, 14]             512
          ReLU-133         [-1, 256, 14, 14]               0
        Conv2d-134         [-1, 256, 14, 14]         589,824
   BatchNorm2d-135         [-1, 256, 14, 14]             512
          ReLU-136         [-1, 256, 14, 14]               0
        Conv2d-137        [-1, 1024, 14, 14]         262,144
   BatchNorm2d-138        [-1, 1024, 14, 14]           2,048
          ReLU-139        [-1, 1024, 14, 14]               0
    Bottleneck-140        [-1, 1024, 14, 14]               0
        Conv2d-141         [-1, 512, 14, 14]         524,288
   BatchNorm2d-142         [-1, 512, 14, 14]           1,024
          ReLU-143         [-1, 512, 14, 14]               0
        Conv2d-144           [-1, 512, 7, 7]       2,359,296
```

```
     BatchNorm2d-145          [-1, 512, 7, 7]           1,024
            ReLU-146          [-1, 512, 7, 7]               0
          Conv2d-147         [-1, 2048, 7, 7]       1,048,576
     BatchNorm2d-148         [-1, 2048, 7, 7]           4,096
          Conv2d-149         [-1, 2048, 7, 7]       2,097,152
     BatchNorm2d-150         [-1, 2048, 7, 7]           4,096
            ReLU-151         [-1, 2048, 7, 7]               0
      Bottleneck-152         [-1, 2048, 7, 7]               0
          Conv2d-153          [-1, 512, 7, 7]       1,048,576
     BatchNorm2d-154          [-1, 512, 7, 7]           1,024
            ReLU-155          [-1, 512, 7, 7]               0
          Conv2d-156          [-1, 512, 7, 7]       2,359,296
     BatchNorm2d-157          [-1, 512, 7, 7]           1,024
            ReLU-158          [-1, 512, 7, 7]               0
          Conv2d-159         [-1, 2048, 7, 7]       1,048,576
     BatchNorm2d-160         [-1, 2048, 7, 7]           4,096
            ReLU-161         [-1, 2048, 7, 7]               0
      Bottleneck-162         [-1, 2048, 7, 7]               0
          Conv2d-163          [-1, 512, 7, 7]       1,048,576
     BatchNorm2d-164          [-1, 512, 7, 7]           1,024
            ReLU-165          [-1, 512, 7, 7]               0
          Conv2d-166          [-1, 512, 7, 7]       2,359,296
     BatchNorm2d-167          [-1, 512, 7, 7]           1,024
            ReLU-168          [-1, 512, 7, 7]               0
          Conv2d-169         [-1, 2048, 7, 7]       1,048,576
     BatchNorm2d-170         [-1, 2048, 7, 7]           4,096
            ReLU-171         [-1, 2048, 7, 7]               0
      Bottleneck-172         [-1, 2048, 7, 7]               0
AdaptiveAvgPool2d-173         [-1, 2048, 1, 1]               0
          Linear-174                   [-1, 3]           6,147
          ResNet-175                   [-1, 3]               0
================================================================
Total params: 23,514,179
Trainable params: 6,147
Non-trainable params: 23,508,032
----------------------------------------------------------------
Input size (MB): 0.57
Forward/backward pass size (MB): 286.55
Params size (MB): 89.70
Estimated Total Size (MB): 376.82
```

**Table [3.3] Proposed Architecture of the ResNet model**

The summary in table [3.3] provides a detailed overview of the ResNet-50 model fine-tuned for a 3-class classification task. The architecture begins with an initial convolutional layer, Conv2d(1), which uses 64 filters with a 7x7 kernel size, stride of 2, and padding of 3, followed by BatchNorm2d, ReLU activation, and MaxPool2d with a 3x3 kernel, stride of 2, and padding of 1. The model comprises bottleneck residual blocks, each containing three Conv2d layers followed by BatchNorm and ReLU, with shortcut connections that may be either identity or projection to match dimensions. The network transitions through stages with increasing depth and feature map dimensions, often involving downsampling via strided convolutions. Following the convolutional layers, an AdaptiveAvgPool2d layer reduces the spatial dimensions to 1x1, and a final Linear layer maps the 2048-dimensional feature vector to 3 output classes. The model contains 23,514,179 total parameters, of which 6,147 are trainable and 23,508,032 are non-trainable, indicating that most parameters are frozen from a pre-trained model. The memory consumption includes an input size of 0.57 MB, forward/backward pass size of 286.55 MB, params size of 89.70 MB, and a total size of 376.82 MB, aligning with typical expectations for deep CNNs like ResNet-50.

### 3.3.3 : GoogLeNet

When layers are increased in neural networks, significant performance gain is achieved. But at the same time extensive networks are prone to overfitting due to exploding or vanishing gradients. This problem was taken care by GoogLeNet architecture [32] which introduced Inception module. Through dimensional reduction, inception module detects features through convolutional layers with different filters applied. GoogLeNet architecture is comprised of 22 layers including 9 inception modules.

In Inception module, 1 x 1, 3 x 3 and 5 x 5 convolutions are used and along with this 3 x 3 max pooling is applied on the previous layer and then their outputs are concatenated and passed to the next layers as shown in Fig. [3.6]. After last inception module, average pooling is applied in order to take average across all feature maps and as a result height and width is reduced to 1 x 1. Moreover this, dropout of 40% is used as a regularization technique to prevent overfitting. In addition to it, Auxiliary classifiers are implemented to intermediated third and sixth inception modules.

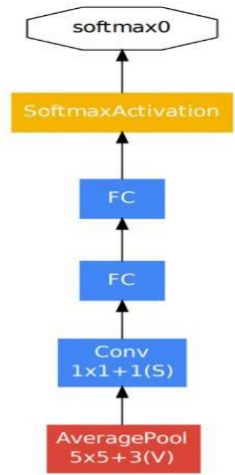

**Fig. [3.6] Auxiliary Classifier in GoogLeNet Architecture**

From previous inception modules, activations are received as input to the Auxiliary classifier and it consists average pooling layers, convolution layers, two fully connected layers, 70% dropout layer followed by linear layer and a SoftMax activation function as shown in Fig. [3.6].

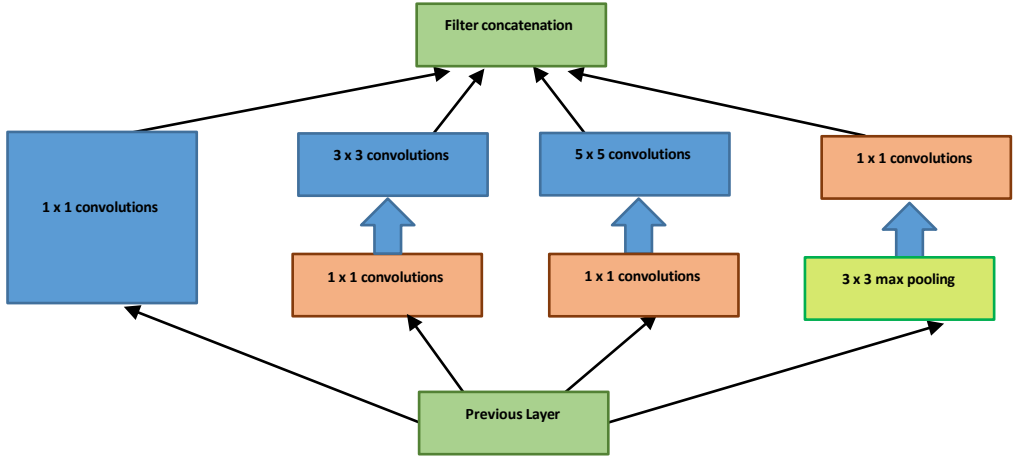

**Fig [3.7] Inception Module in GoogLeNet Architecture**

**Architecture of the GoogLeNet model:**

```
----------------------------------------------------------------
        Layer (type)               Output Shape         Param #
================================================================
            Conv2d-1         [-1, 64, 112, 112]           9,408
       BatchNorm2d-2         [-1, 64, 112, 112]             128
       BasicConv2d-3         [-1, 64, 112, 112]               0
         MaxPool2d-4           [-1, 64, 56, 56]               0
            Conv2d-5           [-1, 64, 56, 56]           4,096
       BatchNorm2d-6           [-1, 64, 56, 56]             128
       BasicConv2d-7           [-1, 64, 56, 56]               0
            Conv2d-8          [-1, 192, 56, 56]         110,592
       BatchNorm2d-9          [-1, 192, 56, 56]             384
      BasicConv2d-10          [-1, 192, 56, 56]               0
        MaxPool2d-11          [-1, 192, 28, 28]               0
           Conv2d-12           [-1, 64, 28, 28]          12,288
      BatchNorm2d-13           [-1, 64, 28, 28]             128
      BasicConv2d-14           [-1, 64, 28, 28]               0
           Conv2d-15           [-1, 96, 28, 28]          18,432
      BatchNorm2d-16           [-1, 96, 28, 28]             192
      BasicConv2d-17           [-1, 96, 28, 28]               0
           Conv2d-18          [-1, 128, 28, 28]         110,592
      BatchNorm2d-19          [-1, 128, 28, 28]             256
      BasicConv2d-20          [-1, 128, 28, 28]               0
           Conv2d-21           [-1, 16, 28, 28]           3,072
      BatchNorm2d-22           [-1, 16, 28, 28]              32
      BasicConv2d-23           [-1, 16, 28, 28]               0
           Conv2d-24           [-1, 32, 28, 28]           4,608
      BatchNorm2d-25           [-1, 32, 28, 28]              64
      BasicConv2d-26           [-1, 32, 28, 28]               0
        MaxPool2d-27          [-1, 192, 28, 28]               0
           Conv2d-28           [-1, 32, 28, 28]           6,144
```

```
     BatchNorm2d-29            [-1, 32, 28, 28]              64
     BasicConv2d-30            [-1, 32, 28, 28]               0
       Inception-31           [-1, 256, 28, 28]               0
          Conv2d-32           [-1, 128, 28, 28]          32,768
     BatchNorm2d-33           [-1, 128, 28, 28]             256
     BasicConv2d-34           [-1, 128, 28, 28]               0
          Conv2d-35           [-1, 128, 28, 28]          32,768
     BatchNorm2d-36           [-1, 128, 28, 28]             256
     BasicConv2d-37           [-1, 128, 28, 28]               0
          Conv2d-38           [-1, 192, 28, 28]         221,184
     BatchNorm2d-39           [-1, 192, 28, 28]             384
     BasicConv2d-40           [-1, 192, 28, 28]               0
          Conv2d-41            [-1, 32, 28, 28]           8,192
     BatchNorm2d-42            [-1, 32, 28, 28]              64
     BasicConv2d-43            [-1, 32, 28, 28]               0
          Conv2d-44            [-1, 96, 28, 28]          27,648
     BatchNorm2d-45            [-1, 96, 28, 28]             192
     BasicConv2d-46            [-1, 96, 28, 28]               0
       MaxPool2d-47           [-1, 256, 28, 28]               0
          Conv2d-48            [-1, 64, 28, 28]          16,384
     BatchNorm2d-49            [-1, 64, 28, 28]             128
     BasicConv2d-50            [-1, 64, 28, 28]               0
       Inception-51           [-1, 480, 28, 28]               0
       MaxPool2d-52           [-1, 480, 14, 14]               0
          Conv2d-53           [-1, 192, 14, 14]          92,160
     BatchNorm2d-54           [-1, 192, 14, 14]             384
     BasicConv2d-55           [-1, 192, 14, 14]               0
          Conv2d-56            [-1, 96, 14, 14]          46,080
     BatchNorm2d-57            [-1, 96, 14, 14]             192
     BasicConv2d-58            [-1, 96, 14, 14]               0
          Conv2d-59           [-1, 208, 14, 14]         179,712
     BatchNorm2d-60           [-1, 208, 14, 14]             416
     BasicConv2d-61           [-1, 208, 14, 14]               0
          Conv2d-62            [-1, 16, 14, 14]           7,680
     BatchNorm2d-63            [-1, 16, 14, 14]              32
     BasicConv2d-64            [-1, 16, 14, 14]               0
          Conv2d-65            [-1, 48, 14, 14]           6,912
     BatchNorm2d-66            [-1, 48, 14, 14]              96
     BasicConv2d-67            [-1, 48, 14, 14]               0
       MaxPool2d-68           [-1, 480, 14, 14]               0
          Conv2d-69            [-1, 64, 14, 14]          30,720
     BatchNorm2d-70            [-1, 64, 14, 14]             128
     BasicConv2d-71            [-1, 64, 14, 14]               0
       Inception-72           [-1, 512, 14, 14]               0
          Conv2d-73           [-1, 160, 14, 14]          81,920
     BatchNorm2d-74           [-1, 160, 14, 14]             320
     BasicConv2d-75           [-1, 160, 14, 14]               0
          Conv2d-76           [-1, 112, 14, 14]          57,344
     BatchNorm2d-77           [-1, 112, 14, 14]             224
     BasicConv2d-78           [-1, 112, 14, 14]               0
          Conv2d-79           [-1, 224, 14, 14]         225,792
     BatchNorm2d-80           [-1, 224, 14, 14]             448
     BasicConv2d-81           [-1, 224, 14, 14]               0
          Conv2d-82            [-1, 24, 14, 14]          12,288
```

```
    BatchNorm2d-83          [-1, 24, 14, 14]              48
    BasicConv2d-84          [-1, 24, 14, 14]               0
         Conv2d-85          [-1, 64, 14, 14]          13,824
    BatchNorm2d-86          [-1, 64, 14, 14]             128
    BasicConv2d-87          [-1, 64, 14, 14]               0
      MaxPool2d-88         [-1, 512, 14, 14]               0
         Conv2d-89          [-1, 64, 14, 14]          32,768
    BatchNorm2d-90          [-1, 64, 14, 14]             128
    BasicConv2d-91          [-1, 64, 14, 14]               0
      Inception-92         [-1, 512, 14, 14]               0
         Conv2d-93         [-1, 128, 14, 14]          65,536
    BatchNorm2d-94         [-1, 128, 14, 14]             256
    BasicConv2d-95         [-1, 128, 14, 14]               0
         Conv2d-96         [-1, 128, 14, 14]          65,536
    BatchNorm2d-97         [-1, 128, 14, 14]             256
    BasicConv2d-98         [-1, 128, 14, 14]               0
         Conv2d-99         [-1, 256, 14, 14]         294,912
   BatchNorm2d-100         [-1, 256, 14, 14]             512
   BasicConv2d-101         [-1, 256, 14, 14]               0
        Conv2d-102          [-1, 24, 14, 14]          12,288
   BatchNorm2d-103          [-1, 24, 14, 14]              48
   BasicConv2d-104          [-1, 24, 14, 14]               0
        Conv2d-105          [-1, 64, 14, 14]          13,824
   BatchNorm2d-106          [-1, 64, 14, 14]             128
   BasicConv2d-107          [-1, 64, 14, 14]               0
     MaxPool2d-108         [-1, 512, 14, 14]               0
        Conv2d-109          [-1, 64, 14, 14]          32,768
   BatchNorm2d-110          [-1, 64, 14, 14]             128
   BasicConv2d-111          [-1, 64, 14, 14]               0
     Inception-112         [-1, 512, 14, 14]               0
        Conv2d-113         [-1, 112, 14, 14]          57,344
   BatchNorm2d-114         [-1, 112, 14, 14]             224
   BasicConv2d-115         [-1, 112, 14, 14]               0
        Conv2d-116         [-1, 144, 14, 14]          73,728
   BatchNorm2d-117         [-1, 144, 14, 14]             288
   BasicConv2d-118         [-1, 144, 14, 14]               0
        Conv2d-119         [-1, 288, 14, 14]         373,248
   BatchNorm2d-120         [-1, 288, 14, 14]             576
   BasicConv2d-121         [-1, 288, 14, 14]               0
        Conv2d-122          [-1, 32, 14, 14]          16,384
   BatchNorm2d-123          [-1, 32, 14, 14]              64
   BasicConv2d-124          [-1, 32, 14, 14]               0
        Conv2d-125          [-1, 64, 14, 14]          18,432
   BatchNorm2d-126          [-1, 64, 14, 14]             128
   BasicConv2d-127          [-1, 64, 14, 14]               0
     MaxPool2d-128         [-1, 512, 14, 14]               0
        Conv2d-129          [-1, 64, 14, 14]          32,768
   BatchNorm2d-130          [-1, 64, 14, 14]             128
   BasicConv2d-131          [-1, 64, 14, 14]               0
     Inception-132         [-1, 528, 14, 14]               0
        Conv2d-133         [-1, 256, 14, 14]         135,168
   BatchNorm2d-134         [-1, 256, 14, 14]             512
   BasicConv2d-135         [-1, 256, 14, 14]               0
        Conv2d-136         [-1, 160, 14, 14]          84,480
```

```
     BatchNorm2d-137          [-1, 160, 14, 14]              320
     BasicConv2d-138          [-1, 160, 14, 14]                0
          Conv2d-139          [-1, 320, 14, 14]          460,800
     BatchNorm2d-140          [-1, 320, 14, 14]              640
     BasicConv2d-141          [-1, 320, 14, 14]                0
          Conv2d-142           [-1, 32, 14, 14]           16,896
     BatchNorm2d-143           [-1, 32, 14, 14]               64
     BasicConv2d-144           [-1, 32, 14, 14]                0
          Conv2d-145          [-1, 128, 14, 14]           36,864
     BatchNorm2d-146          [-1, 128, 14, 14]              256
     BasicConv2d-147          [-1, 128, 14, 14]                0
       MaxPool2d-148          [-1, 528, 14, 14]                0
          Conv2d-149          [-1, 128, 14, 14]           67,584
     BatchNorm2d-150          [-1, 128, 14, 14]              256
     BasicConv2d-151          [-1, 128, 14, 14]                0
       Inception-152          [-1, 832, 14, 14]                0
       MaxPool2d-153            [-1, 832, 7, 7]                0
          Conv2d-154            [-1, 256, 7, 7]          212,992
     BatchNorm2d-155            [-1, 256, 7, 7]              512
     BasicConv2d-156            [-1, 256, 7, 7]                0
          Conv2d-157            [-1, 160, 7, 7]          133,120
     BatchNorm2d-158            [-1, 160, 7, 7]              320
     BasicConv2d-159            [-1, 160, 7, 7]                0
          Conv2d-160            [-1, 320, 7, 7]          460,800
     BatchNorm2d-161            [-1, 320, 7, 7]              640
     BasicConv2d-162            [-1, 320, 7, 7]                0
          Conv2d-163             [-1, 32, 7, 7]           26,624
     BatchNorm2d-164             [-1, 32, 7, 7]               64
     BasicConv2d-165             [-1, 32, 7, 7]                0
          Conv2d-166            [-1, 128, 7, 7]           36,864
     BatchNorm2d-167            [-1, 128, 7, 7]              256
     BasicConv2d-168            [-1, 128, 7, 7]                0
       MaxPool2d-169            [-1, 832, 7, 7]                0
          Conv2d-170            [-1, 128, 7, 7]          106,496
     BatchNorm2d-171            [-1, 128, 7, 7]              256
     BasicConv2d-172            [-1, 128, 7, 7]                0
       Inception-173            [-1, 832, 7, 7]                0
          Conv2d-174            [-1, 384, 7, 7]          319,488
     BatchNorm2d-175            [-1, 384, 7, 7]              768
     BasicConv2d-176            [-1, 384, 7, 7]                0
          Conv2d-177            [-1, 192, 7, 7]          159,744
     BatchNorm2d-178            [-1, 192, 7, 7]              384
     BasicConv2d-179            [-1, 192, 7, 7]                0
          Conv2d-180            [-1, 384, 7, 7]          663,552
     BatchNorm2d-181            [-1, 384, 7, 7]              768
     BasicConv2d-182            [-1, 384, 7, 7]                0
          Conv2d-183             [-1, 48, 7, 7]           39,936
     BatchNorm2d-184             [-1, 48, 7, 7]               96
     BasicConv2d-185             [-1, 48, 7, 7]                0
          Conv2d-186            [-1, 128, 7, 7]           55,296
     BatchNorm2d-187            [-1, 128, 7, 7]              256
     BasicConv2d-188            [-1, 128, 7, 7]                0
       MaxPool2d-189            [-1, 832, 7, 7]                0
          Conv2d-190            [-1, 128, 7, 7]          106,496
```

```
        BatchNorm2d-191           [-1, 128, 7, 7]             256
        BasicConv2d-192           [-1, 128, 7, 7]               0
          Inception-193          [-1, 1024, 7, 7]               0
  AdaptiveAvgPool2d-194          [-1, 1024, 1, 1]               0
            Dropout-195               [-1, 1024]                0
             Linear-196                  [-1, 3]            3,075
          GoogLeNet-197                  [-1, 3]                0
================================================================
Total params: 5,602,979
Trainable params: 3,075
Non-trainable params: 5,599,904
----------------------------------------------------------------
Input size (MB): 0.57
Forward/backward pass size (MB): 94.10
Params size (MB): 21.37
Estimated Total Size (MB): 116.05
```

**Table [3.4] Proposed Architecture of the GoogLeNet model**

The provided summary in table [3.4] describes a deep learning model, specifically a variant of the GoogLeNet architecture, also known as Inception v1. The model structure begins with initial convolutional layers where Conv2d-1 transforms the input using 64 filters of size 7x7, resulting in an output of shape [-1, 64, 112, 112], followed by BatchNorm2d-2 for batch normalization, and combined as BasicConv2d-3. MaxPool2d-4 then reduces the spatial dimensions by half, producing an output of [-1, 64, 56, 56]. The second convolutional block includes Conv2d-5 to BasicConv2d-7 with 64 filters of size 1x1 followed by batch normalization and a convolutional block with 192 filters of size 3x3, with MaxPool2d-11 further reducing the dimensions to [-1, 192, 28, 28]. The core of the network includes multiple inception modules, such as Inception-31, which combines outputs of various convolutions and pooling operations, forming a multi-channel output of shape [-1, 256, 28, 28], capturing features at multiple scales. The reduction and classification phase involves MaxPool2d-52 reducing spatial dimensions to [-1, 480, 14, 14], with subsequent inception modules increasing channel depth to 832 and finally 1024. AdaptiveAvgPool2d-194 then reduces each feature map to a single value (1x1 spatial dimension), followed by Dropout-195 to mitigate overfitting, and Linear-196 which maps the 1024 features to 3 output classes, indicating a classification model with 3 classes. The model comprises 5,602,979 total parameters, with 3,075 trainable parameters (likely only the final classification layer) and 5,599,904 non-trainable parameters (frozen layers, typical in transfer learning). The memory usage includes an input size of 0.57 MB, forward/backward pass size of 94.10 MB, params size of 21.37 MB, and an estimated total size of 116.05 MB.

### 3.3.4 : MNAST:

MNAS Net [36] is a deep learning model designed for image classification tasks on mobile and embedded devices. It utilizes a technique called neural architecture search (NAS) to find an efficient architecture that balances accuracy with computational cost. The model can be used for transfer learning, where a pre-trained version on a large dataset like ImageNet is fine-tuned for a specific image classification task on a new dataset. This approach allows you to leverage the power of a pre-trained model while requiring less computational resources and training data compared to training a model from scratch.

A reinforcement learning method is employed to discover model architectures that maximize the proposed objective function. In each iteration, the controller begins by sampling a batch of models, predicting a sequence of tokens using its RNN network. Each sampled model mmm is then trained on the ImageNet dataset for 5 epochs to determine its accuracy ACC(m), and it is executed on actual phones to measure its latency LAT(m). At the end of the iteration, R(m) is calculated as per the previously described formula, and the RNN parameters are updated using Proximal Policy Optimization (PPO).

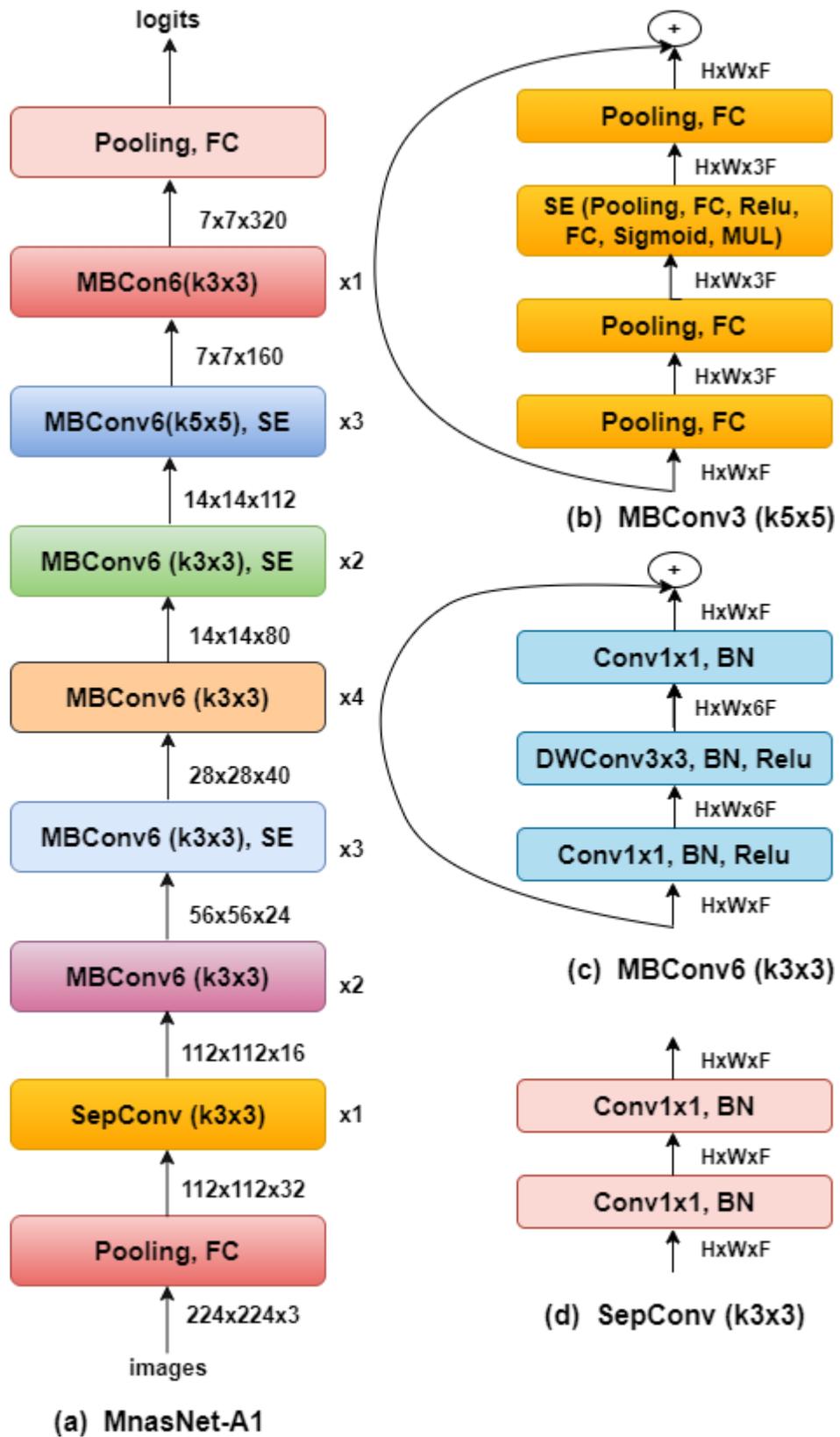

**Fig [3.8] MNAS Net Architecture**

**Architecture of the MNAS Net model:**

```
----------------------------------------------------------------
        Layer (type)               Output Shape         Param #
================================================================
            Conv2d-1         [-1, 32, 112, 112]             864
       BatchNorm2d-2         [-1, 32, 112, 112]              64
              ReLU-3         [-1, 32, 112, 112]               0
            Conv2d-4         [-1, 32, 112, 112]             288
       BatchNorm2d-5         [-1, 32, 112, 112]              64
              ReLU-6         [-1, 32, 112, 112]               0
            Conv2d-7         [-1, 16, 112, 112]             512
       BatchNorm2d-8         [-1, 16, 112, 112]              32
            Conv2d-9         [-1, 48, 112, 112]             768
      BatchNorm2d-10         [-1, 48, 112, 112]              96
             ReLU-11         [-1, 48, 112, 112]               0
           Conv2d-12           [-1, 48, 56, 56]             432
      BatchNorm2d-13           [-1, 48, 56, 56]              96
             ReLU-14           [-1, 48, 56, 56]               0
           Conv2d-15           [-1, 24, 56, 56]           1,152
      BatchNorm2d-16           [-1, 24, 56, 56]              48
_InvertedResidual-17           [-1, 24, 56, 56]               0
           Conv2d-18           [-1, 72, 56, 56]           1,728
      BatchNorm2d-19           [-1, 72, 56, 56]             144
             ReLU-20           [-1, 72, 56, 56]               0
           Conv2d-21           [-1, 72, 56, 56]             648
      BatchNorm2d-22           [-1, 72, 56, 56]             144
             ReLU-23           [-1, 72, 56, 56]               0
           Conv2d-24           [-1, 24, 56, 56]           1,728
      BatchNorm2d-25           [-1, 24, 56, 56]              48
_InvertedResidual-26           [-1, 24, 56, 56]               0
           Conv2d-27           [-1, 72, 56, 56]           1,728
      BatchNorm2d-28           [-1, 72, 56, 56]             144
             ReLU-29           [-1, 72, 56, 56]               0
           Conv2d-30           [-1, 72, 56, 56]             648
      BatchNorm2d-31           [-1, 72, 56, 56]             144
             ReLU-32           [-1, 72, 56, 56]               0
           Conv2d-33           [-1, 24, 56, 56]           1,728
      BatchNorm2d-34           [-1, 24, 56, 56]              48
_InvertedResidual-35           [-1, 24, 56, 56]               0
           Conv2d-36           [-1, 72, 56, 56]           1,728
      BatchNorm2d-37           [-1, 72, 56, 56]             144
             ReLU-38           [-1, 72, 56, 56]               0
           Conv2d-39           [-1, 72, 28, 28]           1,800
      BatchNorm2d-40           [-1, 72, 28, 28]             144
             ReLU-41           [-1, 72, 28, 28]               0
           Conv2d-42           [-1, 40, 28, 28]           2,880
      BatchNorm2d-43           [-1, 40, 28, 28]              80
_InvertedResidual-44           [-1, 40, 28, 28]               0
           Conv2d-45          [-1, 120, 28, 28]           4,800
      BatchNorm2d-46          [-1, 120, 28, 28]             240
             ReLU-47          [-1, 120, 28, 28]               0
           Conv2d-48          [-1, 120, 28, 28]           3,000
      BatchNorm2d-49          [-1, 120, 28, 28]             240
```

```
            ReLU-50          [-1, 120, 28, 28]               0
          Conv2d-51           [-1, 40, 28, 28]           4,800
     BatchNorm2d-52           [-1, 40, 28, 28]              80
_InvertedResidual-53          [-1, 40, 28, 28]               0
          Conv2d-54          [-1, 120, 28, 28]           4,800
     BatchNorm2d-55          [-1, 120, 28, 28]             240
            ReLU-56          [-1, 120, 28, 28]               0
          Conv2d-57          [-1, 120, 28, 28]           3,000
     BatchNorm2d-58          [-1, 120, 28, 28]             240
            ReLU-59          [-1, 120, 28, 28]               0
          Conv2d-60           [-1, 40, 28, 28]           4,800
     BatchNorm2d-61           [-1, 40, 28, 28]              80
_InvertedResidual-62          [-1, 40, 28, 28]               0
          Conv2d-63          [-1, 240, 28, 28]           9,600
     BatchNorm2d-64          [-1, 240, 28, 28]             480
            ReLU-65          [-1, 240, 28, 28]               0
          Conv2d-66          [-1, 240, 14, 14]           6,000
     BatchNorm2d-67          [-1, 240, 14, 14]             480
            ReLU-68          [-1, 240, 14, 14]               0
          Conv2d-69           [-1, 80, 14, 14]          19,200
     BatchNorm2d-70           [-1, 80, 14, 14]             160
_InvertedResidual-71          [-1, 80, 14, 14]               0
          Conv2d-72          [-1, 480, 14, 14]          38,400
     BatchNorm2d-73          [-1, 480, 14, 14]             960
            ReLU-74          [-1, 480, 14, 14]               0
          Conv2d-75          [-1, 480, 14, 14]          12,000
     BatchNorm2d-76          [-1, 480, 14, 14]             960
            ReLU-77          [-1, 480, 14, 14]               0
          Conv2d-78           [-1, 80, 14, 14]          38,400
     BatchNorm2d-79           [-1, 80, 14, 14]             160
_InvertedResidual-80          [-1, 80, 14, 14]               0
          Conv2d-81          [-1, 480, 14, 14]          38,400
     BatchNorm2d-82          [-1, 480, 14, 14]             960
            ReLU-83          [-1, 480, 14, 14]               0
          Conv2d-84          [-1, 480, 14, 14]          12,000
     BatchNorm2d-85          [-1, 480, 14, 14]             960
            ReLU-86          [-1, 480, 14, 14]               0
          Conv2d-87           [-1, 80, 14, 14]          38,400
     BatchNorm2d-88           [-1, 80, 14, 14]             160
_InvertedResidual-89          [-1, 80, 14, 14]               0
          Conv2d-90          [-1, 480, 14, 14]          38,400
     BatchNorm2d-91          [-1, 480, 14, 14]             960
            ReLU-92          [-1, 480, 14, 14]               0
          Conv2d-93          [-1, 480, 14, 14]           4,320
     BatchNorm2d-94          [-1, 480, 14, 14]             960
            ReLU-95          [-1, 480, 14, 14]               0
          Conv2d-96           [-1, 96, 14, 14]          46,080
     BatchNorm2d-97           [-1, 96, 14, 14]             192
_InvertedResidual-98          [-1, 96, 14, 14]               0
          Conv2d-99          [-1, 576, 14, 14]          55,296
    BatchNorm2d-100          [-1, 576, 14, 14]           1,152
           ReLU-101          [-1, 576, 14, 14]               0
         Conv2d-102          [-1, 576, 14, 14]           5,184
    BatchNorm2d-103          [-1, 576, 14, 14]           1,152
```

```
           ReLU-104          [-1, 576, 14, 14]               0
         Conv2d-105           [-1, 96, 14, 14]          55,296
    BatchNorm2d-106           [-1, 96, 14, 14]             192
_InvertedResidual-107         [-1, 96, 14, 14]               0
         Conv2d-108          [-1, 576, 14, 14]          55,296
    BatchNorm2d-109          [-1, 576, 14, 14]           1,152
           ReLU-110          [-1, 576, 14, 14]               0
         Conv2d-111            [-1, 576, 7, 7]          14,400
    BatchNorm2d-112            [-1, 576, 7, 7]           1,152
           ReLU-113            [-1, 576, 7, 7]               0
         Conv2d-114            [-1, 192, 7, 7]         110,592
    BatchNorm2d-115            [-1, 192, 7, 7]             384
_InvertedResidual-116          [-1, 192, 7, 7]               0
         Conv2d-117           [-1, 1152, 7, 7]         221,184
    BatchNorm2d-118           [-1, 1152, 7, 7]           2,304
           ReLU-119           [-1, 1152, 7, 7]               0
         Conv2d-120           [-1, 1152, 7, 7]          28,800
    BatchNorm2d-121           [-1, 1152, 7, 7]           2,304
           ReLU-122           [-1, 1152, 7, 7]               0
         Conv2d-123            [-1, 192, 7, 7]         221,184
    BatchNorm2d-124            [-1, 192, 7, 7]             384
_InvertedResidual-125          [-1, 192, 7, 7]               0
         Conv2d-126           [-1, 1152, 7, 7]         221,184
    BatchNorm2d-127           [-1, 1152, 7, 7]           2,304
           ReLU-128           [-1, 1152, 7, 7]               0
         Conv2d-129           [-1, 1152, 7, 7]          28,800
    BatchNorm2d-130           [-1, 1152, 7, 7]           2,304
           ReLU-131           [-1, 1152, 7, 7]               0
         Conv2d-132            [-1, 192, 7, 7]         221,184
    BatchNorm2d-133            [-1, 192, 7, 7]             384
_InvertedResidual-134          [-1, 192, 7, 7]               0
         Conv2d-135           [-1, 1152, 7, 7]         221,184
    BatchNorm2d-136           [-1, 1152, 7, 7]           2,304
           ReLU-137           [-1, 1152, 7, 7]               0
         Conv2d-138           [-1, 1152, 7, 7]          28,800
    BatchNorm2d-139           [-1, 1152, 7, 7]           2,304
           ReLU-140           [-1, 1152, 7, 7]               0
         Conv2d-141            [-1, 192, 7, 7]         221,184
    BatchNorm2d-142            [-1, 192, 7, 7]             384
_InvertedResidual-143          [-1, 192, 7, 7]               0
         Conv2d-144           [-1, 1152, 7, 7]         221,184
    BatchNorm2d-145           [-1, 1152, 7, 7]           2,304
           ReLU-146           [-1, 1152, 7, 7]               0
         Conv2d-147           [-1, 1152, 7, 7]          10,368
    BatchNorm2d-148           [-1, 1152, 7, 7]           2,304
           ReLU-149           [-1, 1152, 7, 7]               0
         Conv2d-150            [-1, 320, 7, 7]         368,640
    BatchNorm2d-151            [-1, 320, 7, 7]             640
_InvertedResidual-152          [-1, 320, 7, 7]               0
         Conv2d-153           [-1, 1280, 7, 7]         409,600
    BatchNorm2d-154           [-1, 1280, 7, 7]           2,560
           ReLU-155           [-1, 1280, 7, 7]               0
        Dropout-156                  [-1, 1280]               0
         Linear-157                     [-1, 3]           3,843
```

```
         MNASNet-158                         [-1, 3]                      0
================================================================
Total params: 3,106,155
Trainable params: 3,843
Non-trainable params: 3,102,312
----------------------------------------------------------------------------------------
Input size (MB): 0.57
Forward/backward pass size (MB): 123.37
Params size (MB): 11.85
Estimated Total Size (MB): 135.79
```

**Table [3.5] Proposed Architecture of MNAS Net model**

The model [ 3.5] comprises several key layers, each serving specific functions essential for its performance. The convolutional layers (e.g., Conv2d) perform convolution operations to extract features from input images, with initial layers (e.g., Conv2d-1, Conv2d-4, Conv2d-7) followed by Batch Normalization and ReLU activation functions. Batch Normalization layers (e.g., BatchNorm2d-2, BatchNorm2d-5) stabilize and accelerate training by normalizing inputs to subsequent layers. ReLU activations (e.g., ReLU-3, ReLU-6) introduce non-linearity, enabling the model to learn complex patterns. The model incorporates Inverted Residual Blocks (e.g., _InvertedResidual-17), which consist of convolutions including depth wise separable and pointwise convolutions, reducing parameters while maintaining performance. Depth wise separable convolutions (e.g., Conv2d-12, Conv2d-15) within these blocks further reduce computational cost and model size. Bottleneck layers (e.g., Conv2d-9, Conv2d-54) compress features before expanding them in subsequent layers to minimize parameters. The final layers (e.g., Conv2d-153, Linear-157) transform features into the required output shape for a classification task with three classes. The model has a total of 3,106,155 parameters, with 3,843 trainable and 3,102,312 non-trainable parameters, resulting in an estimated total size of 135.79 MB. The small number of trainable parameters suggests that most of the model's parameters are frozen, likely utilizing a pre-trained model with only the final classification layer being trained for the specific task.

## 3.4 : Experimental set up

In this research work, the experiments are performed on Intel® core ™ i9-10850K at CPU speed 3.6 GHz with 64 GB RAM including an NVIDIA 3060 GPU with 12 GB memory. Operating System was Windows 10 Pro and system type is 64-bit operating system. The summary of experimental configuration is shown in Table [3.6].

| **Operating System** | Windows |
|---|---|
| **CPU** | **Intel® core ™ i9-10850K @ 3.6 GHz, 10 cores, 20 logical processors** |
| **GPU** | NVIDIA GeForce RTX - 3060 |
| **IDE** | **Jupyter Notebook** |
| **Python** | 3.8 |
| **Pytorch** | 1.9.1 |

**Table [3.6] Summary of System Configuration**

This research study is conducted in two phases. In first phase, experimental setup is conducted for generation of images to prepare dataset while in second phase, generated images are passed as input images to proposed classifiers which further identifies images as COVID-19, NORMAL and Viral Pneumonia.

While building real world machine learning models, it is quite common to split the dataset into 3 parts:

1. **Training set** - used to train the model i.e., compute the loss and adjust the weights of the model using gradient descent.
2. **Validation set** - used to evaluate the model while training, adjust hyperparameters (learning rate etc.) and pick the best version of the model.
3. **Test set** - used to compare different models, or different types of modeling approaches, and report the final accuracy of the model.

Since there's no predefined validation set, we set aside a small portion (20 percent) of the training set to be used as the validation set. We'll use the random_split helper method from PyTorch to do this.
Table [3.6.1] shows count of fake images which will be further provided as input to classifier. As shown in Fig [3.9] uniform distribution of images per class is set for training phase while varied distribution is selected for testing purpose to ensure robustness of designed classification model.

| **Dataset/Classes** | **COVID -19** | **Normal** | **Viral Pneumonia** | **Total** |
|---|---|---|---|---|
| **Train** | 4000 | 4000 | 4000 | **12000** |
| **Test** | 59 | 164 | 152 | **375** |
| **Total** | **4059** | **4164** | **4152** | **12375** |

**Table [3.6.1] Expanded number of fake images for each class**

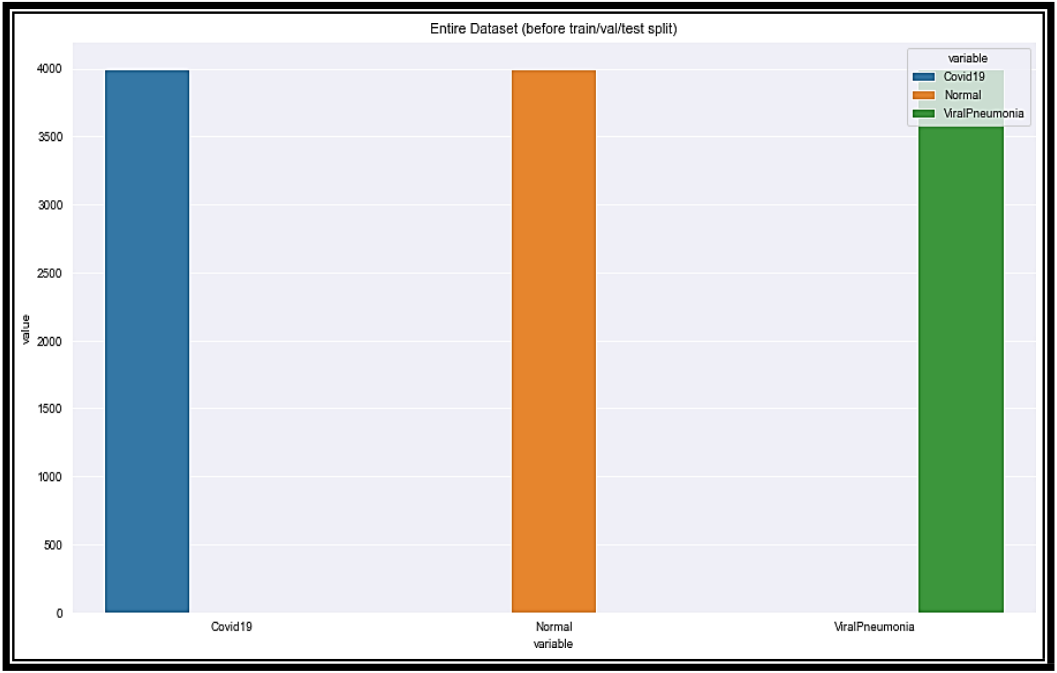

**Fig. [3.9] Class distribution**

### 3.5 : Training of proposed WGAN - GP model

Total 438 Chest X-Ray images are used for training the model of WGAN – GP. To generate images for every class i.e., COVID-19, Normal and Viral Pneumonia., model is trained for every set of images separately.
Step by step training procedure is defined as follows:

- In first step, load all the libraries of Pytorch , Torchvision, Matplotlib and Seaborn.
- Load the dataset of Chest X-ray images.
- Prepare the dataset by converting images in to tensors to work in Pytorch. Resize all the images to 128 * 128 dimensions and normalize all the images within the range of [-1,1].
- Define a function to keep track of gradients.
- Build the neural network of Generator by taking input dimension of noise vector, number of channels and hidden dimension.
- Define the function for creating vectors by giving the dimensions of samples and noise vectors.
- Build the neural network of Critic by taking input as number of channels and hidden dimensions.
- Set all the training initializations as define in table [3.7].
- Initialize weights randomly from a normal distribution with mean = 0 and standard deviation = 0.02. This "weight_init" function takes an initialized model as input and reinitializes all convolutional, transpose convolutional and batch normalization layers to meet these criteria of weight initialization.
- Calculate the gradient penalty in two steps:
    - Compute the gradient with respect to the images and
    - Compute the gradient penalty given the gradient

- Loss of generator and critic is calculated. To calculate the loss of generator, maximize the distance between critic's predictions on real and fake images and add gradient penalty. The gradient penalty is weighed according to λ which is defined in parametric configuration Table [3.7].
- Define directory to save generated images.
- For every generator's update, critic gets updated multiple times and this logic hampers generator to overrule the critic.
- Since this training is executed in multiple shifts so we had to save the model state and then later resumed training by loaded the model.
- Total 2000 epochs are executed to get a reasonable quality of images per class.
- From generated images, best quality 4000 images are selected to pass in to classifier.

### 3.5.1 : Parametric configuration of proposed WGAN - GP model

| **Epochs** | 2000 |
|---|---|
| **Batch Size** | 20 |
| **Optimizer** | Adam [33] |
| **Adam Hyper parameters ($\beta_1, \beta_2$)** | 0.5, 0.0009 [33] |
| **Generator's Learning rate** | 0.0002 |
| **Discriminator's Learning rate** | 0.0002 |
| **Dimension of noise vector(z)** | 128 |
| **Weight of gradient penalty($\lambda c$)** | 10 [33] |
| **No. of critic repeats per generation iteration ($n$critic)** | 5 [33] |

Table [3.7] Parametric Configuration of proposed Classification Models

### 3.6 : Training of proposed Pretrained models

Total 12000 Chest X-Ray images are used for training pre-trained models and 375 images are used to test the model.

Step by step training procedure is defined as follows:

- In first step, load all the libraries of Pytorch, Torchvision, Matplotlib and Seaborn.
- Load the dataset of Chest X-ray images.
- Prepare the dataset by converting images in to tensors to work in Pytorch. Resize all the images to 224 * 224 dimensions and normalize all the images within the range of [-1,1].
- Apply Data augmentation techniques of Random Horizontal Flip, Padding, Random Crop.
- Download the pretrained model from pytorch.org
- Set "feature_extract = true" and update only final layer from which predictions are derived.
- Initialize the model and reshape the output layer according to the number of classes of new dataset.

- Train the model using parametric configuration as shown in table [3.8] and evaluate its performance on the validation set.
- Calculate accuracy and loss of training and validation data.
- Test the model on generated images.

### 3.6.1. Parametric configuration of proposed classification model

|  | VGG -16 | ResNet-50 | GoogLeNet | MNAST |
|---|---|---|---|---|
| **Epochs** | 10 | 30 | 10 | 10 |
| **Batch Size** | 50 | 50 | 50 | 50 |
| **Optimizer** | Adam | Adam | Adam | Adam |
| **Loss Function** | Cross Entropy Loss | Cross Entropy Loss | Cross Entropy Loss | Cross Entropy Loss |
| **Learning rate** | 0.0001 | 0.0001 | 0.0001 | 0.0001 |

Table [3.8] Parametric Configuration of the proposed Classification Models

# Section 4: Results

## 4.1. Wasserstein Loss Function graph

Since WGAN - GP is comprised of two neural networks so for this reason these two networks have their separate loss functions. In WGAN – GP, Wasserstein loss function is employed which influences the critic to give non probabilistic score instead of a probabilistic score of 0 or 1. While training the critic network absolute difference between critic and generator gets maximized. According to Fig [4.1], during generator training, absolute value of the generator function also gets maximized.

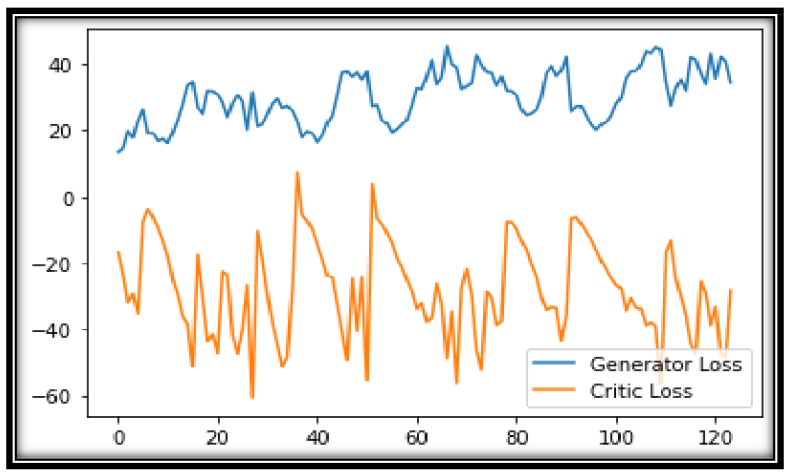

Fig. [4.1] Generator and Critic Loss

## 4.2. Generated Images from WGAN - GP

Our proposed WGAN – GP model generated 40000 images including initial noise images for every class. For classification purpose, we selected best 4000 images from later epochs for each class as shown in Fig. [4.3] (a), (b), (c).

| | |
|---|---|
| **Epoch 1** | 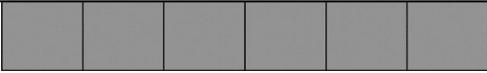 |
| **Epoch 25** | 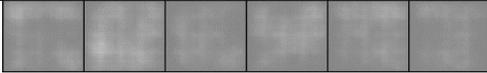 |
| **Epoch 50** | 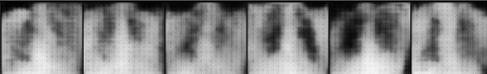 |
| **Epoch 100** | 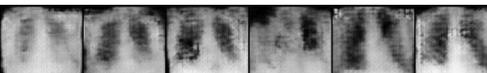 |
| **Epoch 500** | 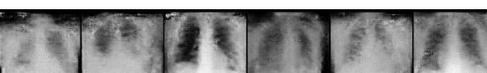 |
| **Epoch 1000** | 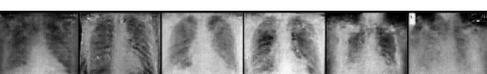 |
| **Epoch 1500** | 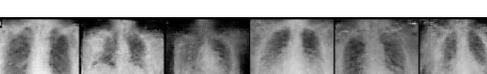 |
| **Epoch 2000** | 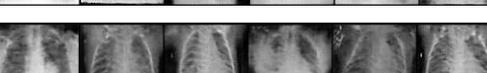 |

**Fig (a) Epoch by Epoch Covid-19 Image generation using WGAN**

| Epoch 1 | 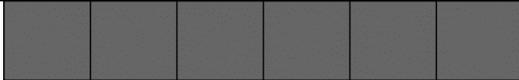 |
| --- | --- |
| Epoch 25 | 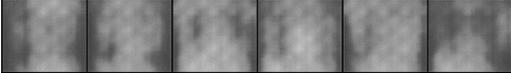 |
| Epoch 50 | 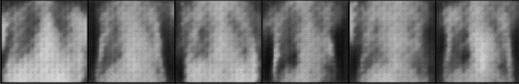 |
| Epoch 100 | 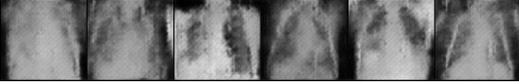 |
| Epoch 500 | 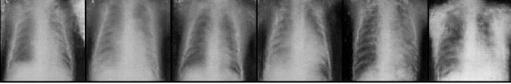 |
| Epoch 1000 | 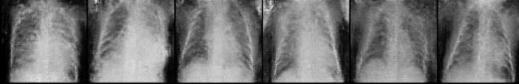 |
| Epoch 1500 | 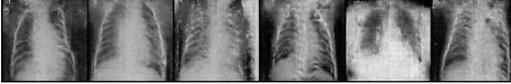 |
| Epoch 2000 | 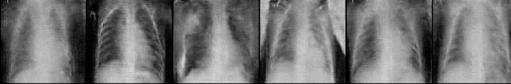 |

**Fig (b) Epoch by Epoch Viral Pneumonia Image generation using WGAN**

| | |
|---|---|
| Epoch 1 | 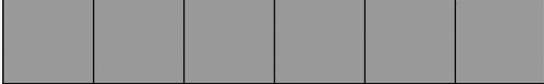 |
| Epoch 25 | 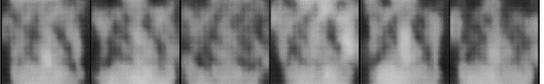 |
| Epoch 50 | 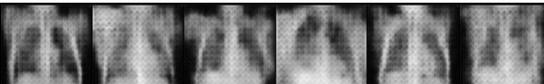 |
| Epoch 100 | 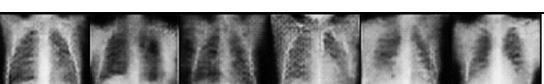 |
| Epoch 500 | 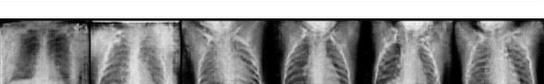 |
| Epoch 1000 | 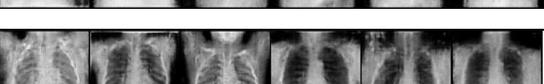 |
| Epoch 1500 | 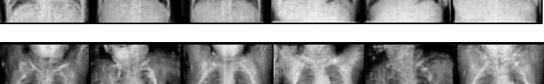 |
| Epoch 2000 | 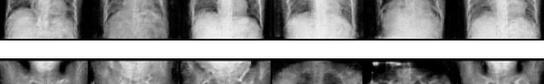 |

**Fig (c) Epoch by Epoch Normal Image generation using WGAN**

**Fig [4.2] Epoch by Epoch Generation of images**

## 4.3 : Results of Classification models

This section presents a comprehensive experimental analysis of COVID-19 prediction, involving 4 different network architectures. We examine the impacts of training and validation losses and accuracies and perform a comparative analysis of the models. Learning curves serve as a widely used tool in diagnosing model's behavior[15-22]. During training, after each update plot of losses and accuracies are created to show learning curves. Evaluation of model on training dataset defines how well model is learning while hold out validation dataset is used to get an idea how well the model is generalizing. To diagnose the behavior of model, Fig. [4.3] shows dual curves of losses and accuracies formed for both validation and training datasets.

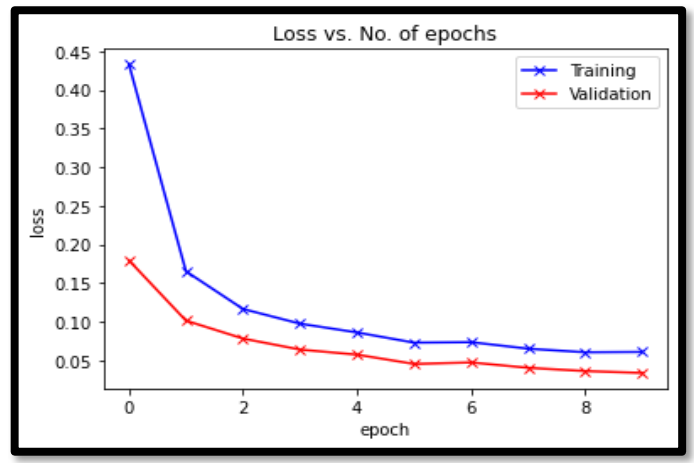

**Graph a -Training Loss and Validation Loss of VGG - 16 model**

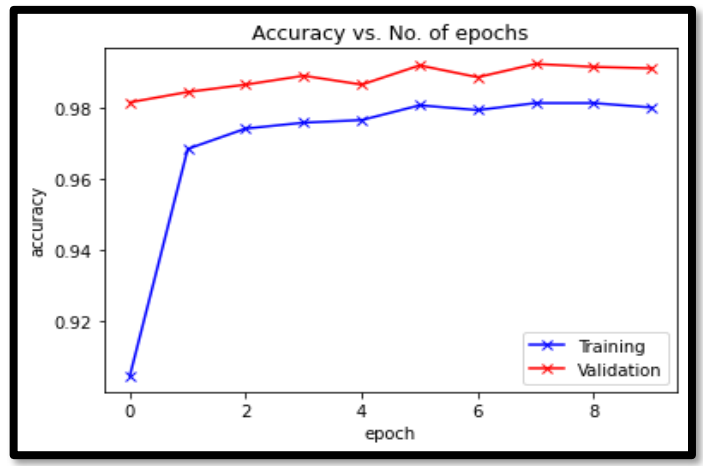

**Graph b - Training Accuracy and Validation Accuracy of VGG -16**

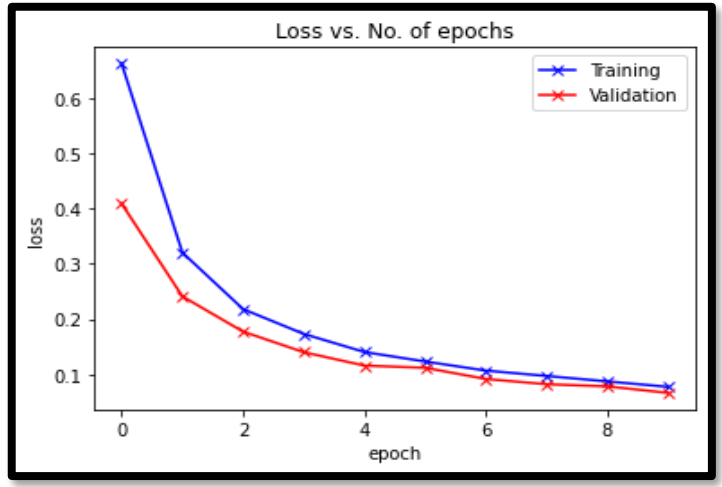

**Graph c -Training Loss and Validation Loss of ResNet-50 model**

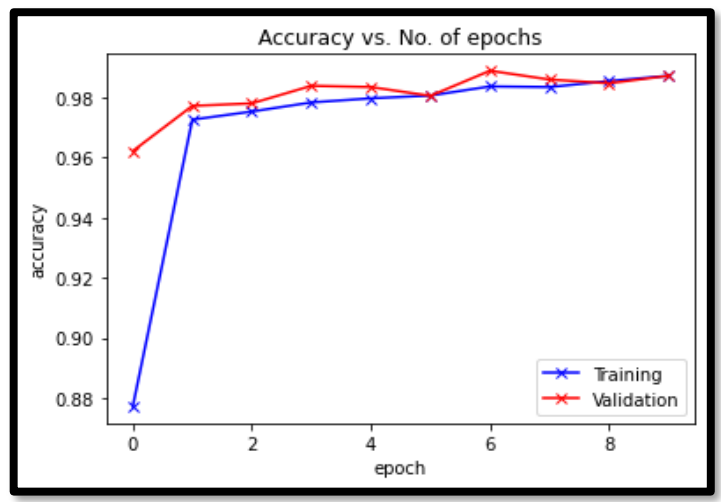

**Graph d - Training Accuracy and Validation Accuracy of ResNet-50 model**

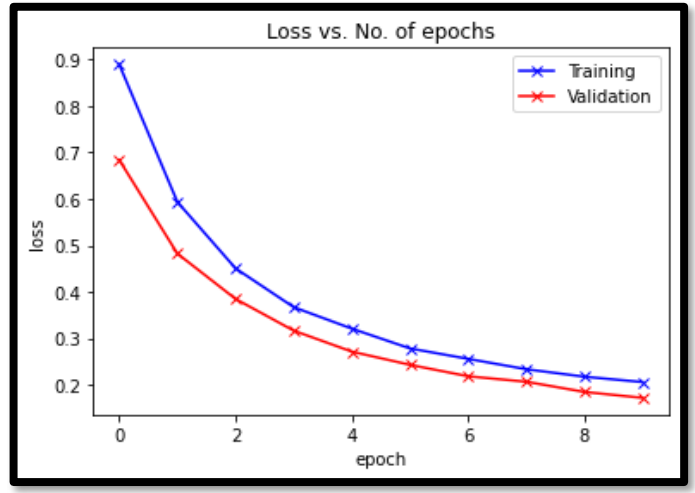

**Graph e -Training Loss and Validation Loss of GoogLeNet model**

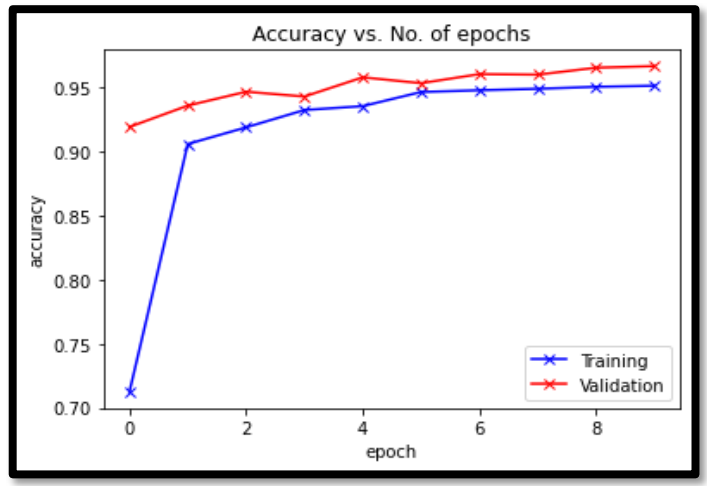

**Graph f - Training Accuracy and Validation Accuracy of GoogLeNet model**

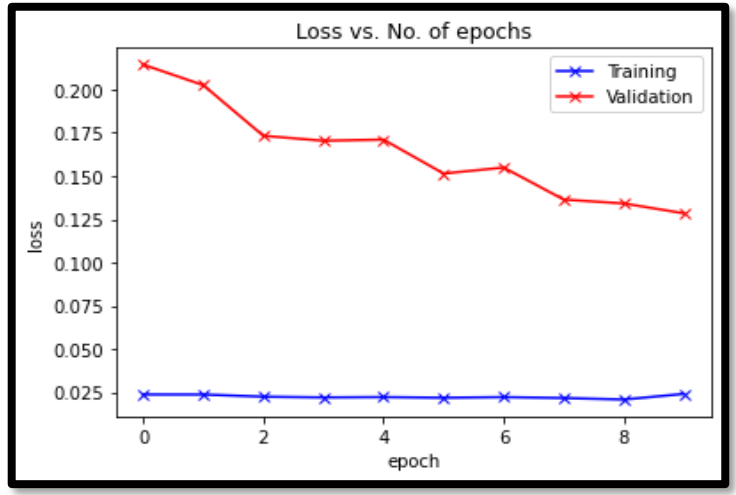

**Graph g - Training Loss and Validation Loss of MNAST Net model**

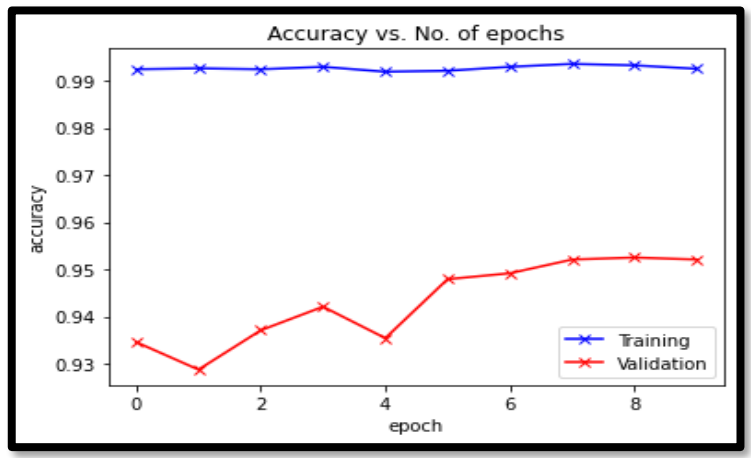

**Graph h - Training Accuracy and Validation Accuracy of MNAST Net model**

**Fig [4.3] Training - Validation Loss and accuracies of classification models**

Graph (a-h) depicts axes labels as epochs and loss where epochs represent number of epochs, or iterations, over which the model is trained and loss shows a measure of how well the model is performing. The pattern observed in the graph(a,c and e) indicate effective learning and generalization of the model. Initially, both training and validation losses are high, which is expected as the model starts with random weights and performs poorly. As training progresses, the losses decrease, showing that the model improves its performance on both datasets. The convergence of training and validation loss lines suggests that the model is learning effectively without overfitting. If the validation loss were to increase while the training loss continued to decrease, it would indicate overfitting. However, in these cases, the convergence of the lines signify good generalization. The optimal point for training is achieved where the validation loss stopped decreasing and stabilized, ensuring the model did not overfit while achieving good performance. Further graph (b,d and f) indicate that both training and validation accuracies started low, which is expected as the models begin with random weights and performs poorly initially. As training progresses, the accuracies increase, demonstrating that the model is improving its performance on both datasets. The training and validation accuracy lines remain parallel and move upwards, suggesting that the models are learning effectively and consistently enhancing its performance. The parallel nature of the lines indicates a constant gap between training and validation accuracy, which is a positive sign that the model is not overfitting. In an overfitting scenario, the training accuracy would continue to improve while the validation accuracy would plateau or decrease. Here in these cases, the consistent gap between the two lines or overlapping shows that the model maintains a similar level of performance on both datasets, ensuring good generalization. The graphs (g and h) of the MNAST model, showing both loss and accuracy, reveal that the lines for training and validation are far apart, indicating a significant discrepancy between the model's performance on the training data versus the validation data. In the loss graph, the large gap suggests that while the model is achieving lower loss (better performance) on the training data, it struggles to generalize well to the validation data, which is indicative of overfitting. Similarly, in the accuracy graph, the training accuracy is much higher than the validation accuracy, reinforcing the idea that the model performs well on the training set but poorly on the validation set. This disparity implies that the model has memorized the training data rather than learning to generalize from it. To address this issue, techniques such as regularization, dropout, or using more training data may be necessary to improve the model's generalization capability and reduce overfitting. Table[4.1] shows training, validation and testing accuracies of all four proposed models.

| Classification Model | Training Accuracy | Testing Accuracy | Validation Accuracy |
|---|---|---|---|
| VGG -16 | 98.02% | 99% | 99.13% |
| ResNet-50 | 99.49% | 93.9% | 99.46% |
| GoogLeNet | 95.14% | 94.49% | 96.67% |
| MNAST Net | 99.25% | 97.75% | 95.21% |

Table [4.1] Parametric Configuration of the proposed Classification Models

## 4.4 : Evaluating Classification Models Using Confusion Matrix

Given a classifier, the best way to evaluate classifier's performance is by using confusion matrix. It is a performance measurement for machine learning classification. It's a table with four different combinations of predicted and actual values.

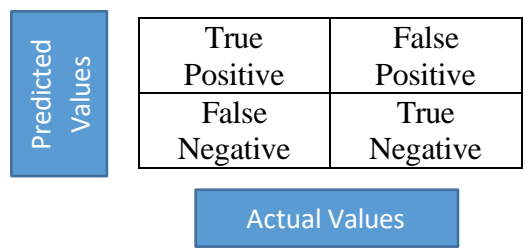

Fig [4.3] Conceptual Diagram of Confusion Matrix

Every time our classifier makes a prediction, one of the cells in the table is incremented by 1. It is extremely useful for measuring Recall, Precision, Accuracy and AUC-ROC curves.

According to Fig. [4.3]

**True Positive (TP):** Model predicted positive and its true.

**True Negative (TN):** Model predicted negative and its true.

**False Positive (FP):** Model predicted positive but its false.

**False Negative (FN):** Model predicted negative and its false.

In addition to above, classification report is a convenient way to calculate precision, recall and F1 score for each class in a multi-class classifier.

**Precision:** Defines what proportion of predicted positive is truly positives.

Mathematically it is expressed as:

$$\frac{TP}{TP + FP}$$

**Recall:** Defines what proportion of actual positives is correctly classified.

Mathematically it is expressed as:

$$\frac{TP}{TP + FN}$$

**Accuracy:** Defines what proportion of both positive and negative are correctly classified.

Mathematically it is expressed as:

$$\frac{TP + TN}{TP + TN + FP + FN}$$

**F1 Score:** Defines performance of a model on the basis of precision and recall. It reaches its best value at 1 and worst value at 0. It uses the harmonic mean which is given by this simple formula:

**F1-score = 2 × (precision × recall)/ (precision + recall)**

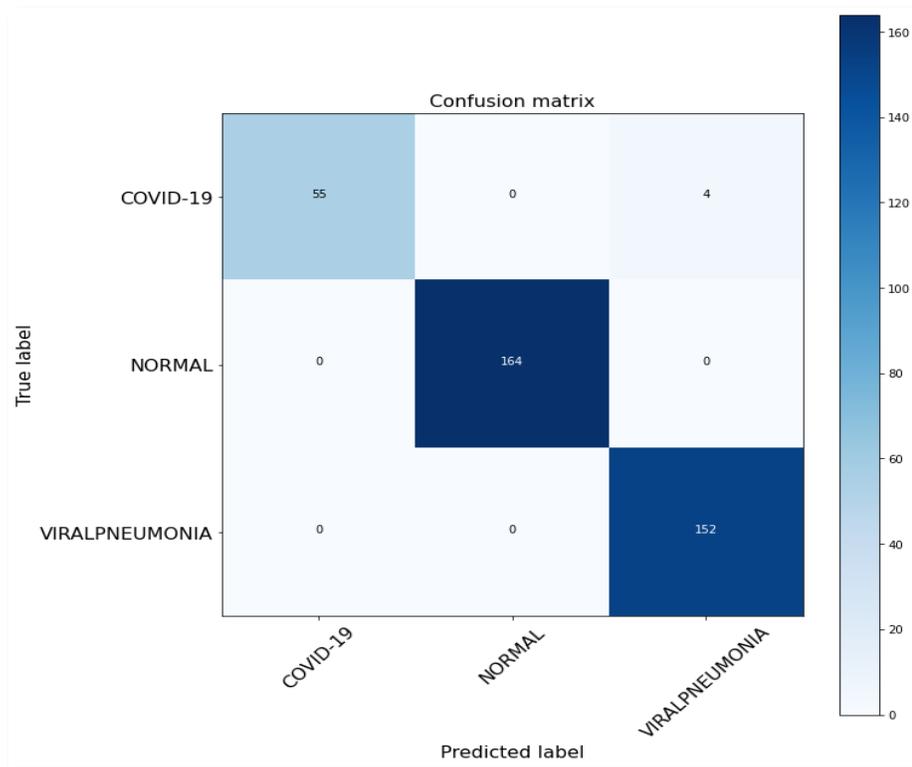

Fig [4.4] Confusion Matrix of VGG -16 Model

```
                precision    recall  f1-score   support

      COVID-19       1.00      0.93      0.96        59
        NORMAL       1.00      1.00      1.00       164
VIRALPNEUMONIA       0.97      1.00      0.99       152

      accuracy                           0.99       375
     macro avg       0.99      0.98      0.98       375
  weighted avg       0.99      0.99      0.99       375
```

Table [4.2] Classification Report of VGG -16 Model

Fig. [4.4] shows confusion matrix for VGG-16 classifier where the model demonstrates outstanding performance across all classes, exhibiting no false positives for any class. However, a small number of false negatives (4) in the COVID-19 class suggest instances where the model misclassified some cases as other classes. Despite this, the model achieves flawless precision, recall, and F1-score for the NORMAL and VIRAL PNEUMONIA classes, indicating its high accuracy in identifying these conditions. .

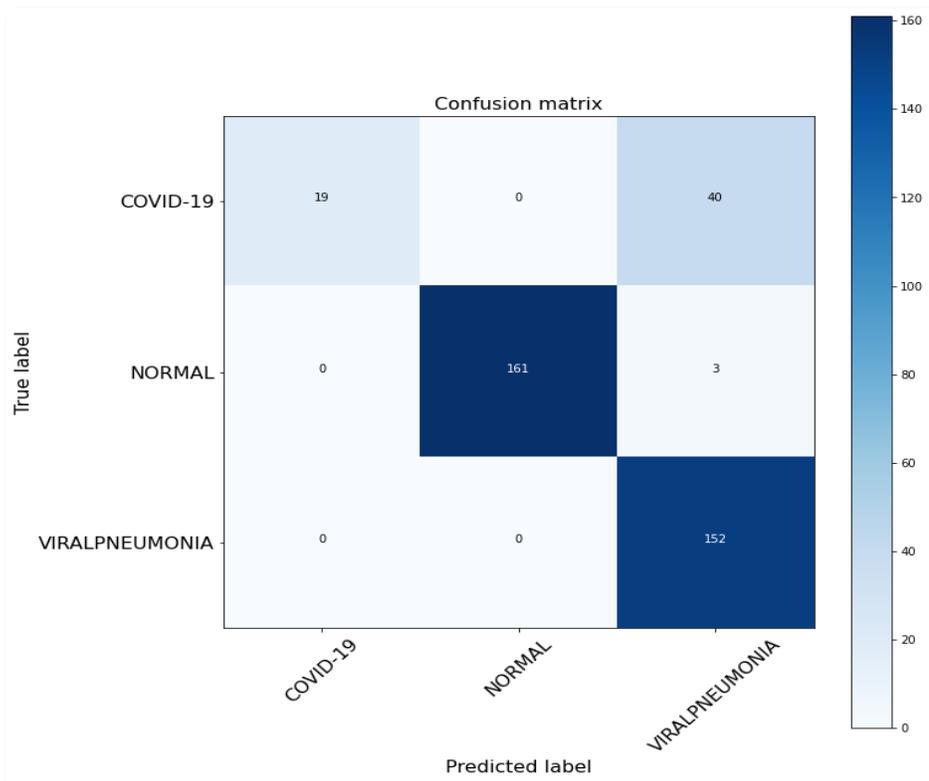

**Fig [4.5] Confusion Matrix of ResNet-50 Model**

The ResNet-50 model in fig. [4.5] exhibits strong performance in accurately identifying instances of VIRAL PNEUMONIA, achieving flawless precision, recall, and F1-score for this class. However, it faces challenges in correctly classifying instances of COVID-19, as evidenced by the relatively high number of false negatives (40). This suggests that the model may struggle to capture the distinctive features of COVID-19 cases compared to other classes. Nonetheless, the false negatives for the NORMAL class are comparatively lower (3), indicating better performance in distinguishing NORMAL cases. Notably, akin to the VGG-16 model, the ResNet-50 model demonstrates zero occurrences of false positives, underscoring its high specificity across all classes.

```
                 precision    recall  f1-score   support

      COVID-19       1.00      0.32      0.49        59
        NORMAL       1.00      0.98      0.99       164
VIRALPNEUMONIA       0.78      1.00      0.88       152

      accuracy                           0.89       375
     macro avg       0.93      0.77      0.78       375
  weighted avg       0.91      0.89      0.87       375
```

**Table [4.3] Classification Report of ResNet -50 Model**

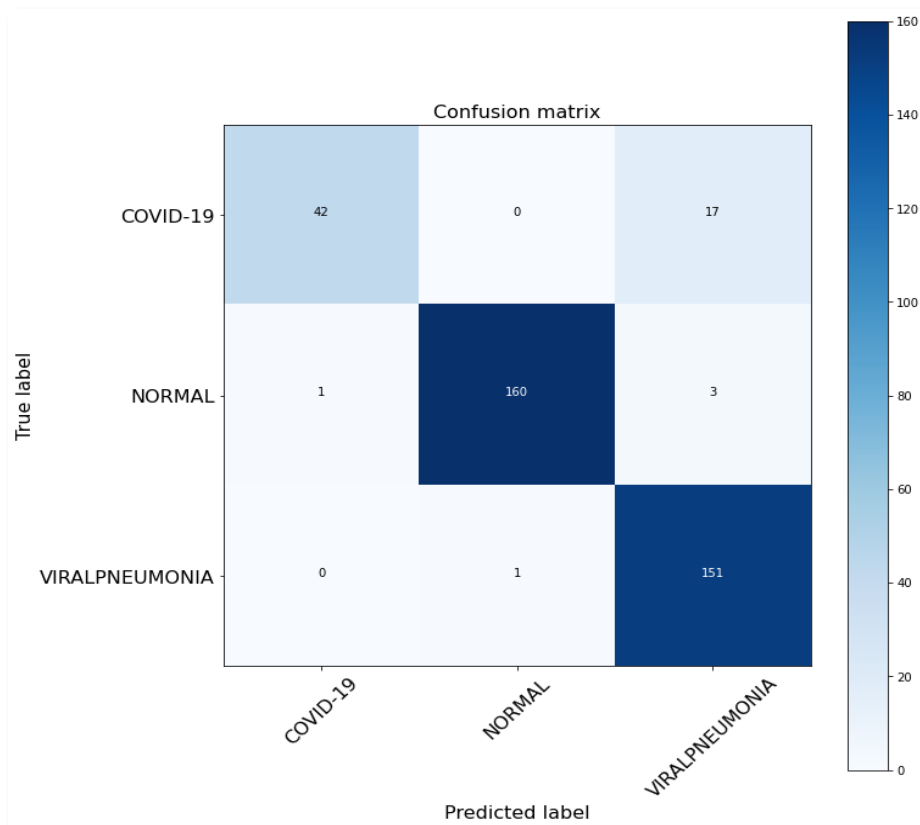

**Fig [4.6] Confusion Matrix of GoogLeNet Model**

```
              precision    recall  f1-score   support

    COVID-19       0.98      0.71      0.82        59
      NORMAL       0.99      0.98      0.98       164
VIRALPNEUMONIA     0.88      0.99      0.93       152

    accuracy                           0.94       375
   macro avg       0.95      0.89      0.91       375
weighted avg       0.95      0.94      0.94       375
```

**Table [4.4] Classification Report of GoogLeNet Model**

Notably, confusion matrix in fig. [4.6] defines that the model excels in accurately classifying instances of NORMAL and VIRAL PNEUMONIA, achieving flawless precision, recall, and F1-score for these classes. This suggests that the model effectively captures the distinguishing features of these conditions, leading to highly accurate predictions. However, the model encounters challenges in correctly identifying instances of COVID-19, as evidenced by the relatively high number of false negatives.

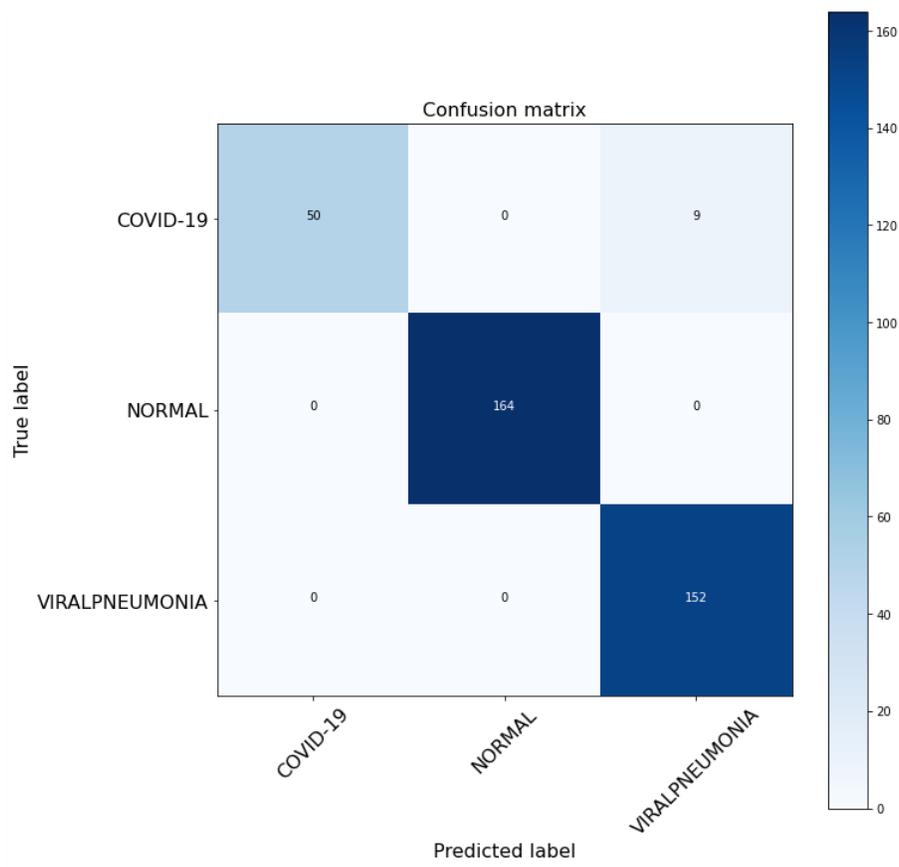

**Fig [4.7] Confusion Matrix of MNAST Model**

```
              precision    recall  f1-score   support

    COVID-19       1.00      0.85      0.92        59
      NORMAL       1.00      1.00      1.00       164
VIRALPNEUMONIA     0.94      1.00      0.97       152

    accuracy                           0.98       375
   macro avg       0.98      0.95      0.96       375
weighted avg       0.98      0.98      0.98       375
```

**Table [4.5] Classification Report of MNAST Model**

The MNAST model demonstrates excellent performance across all classes, achieving perfect precision, recall, and F1-score for COVID-19, NORMAL, and VIRAL PNEUMONIA. Notably, there are no occurrences of false positives or false negatives in the confusion matrix fig[4.7], indicating that the model accurately distinguishes between the three classes without making any misclassifications. The high accuracy and specificity observed in the confusion matrix suggest that the model effectively captures the distinguishing features of each class, leading to reliable predictions.

### 4.5 : Evaluating Classification Models Using Receiver Operating Curve (ROC)

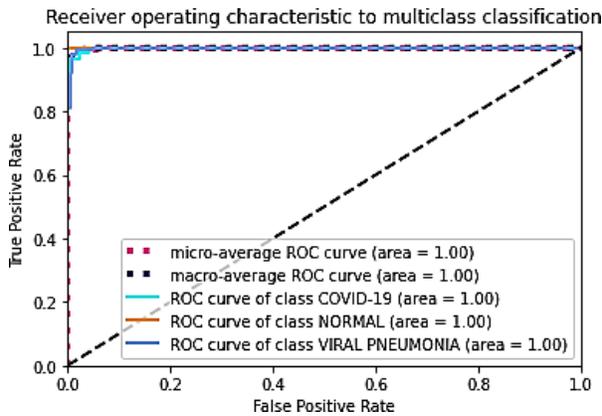
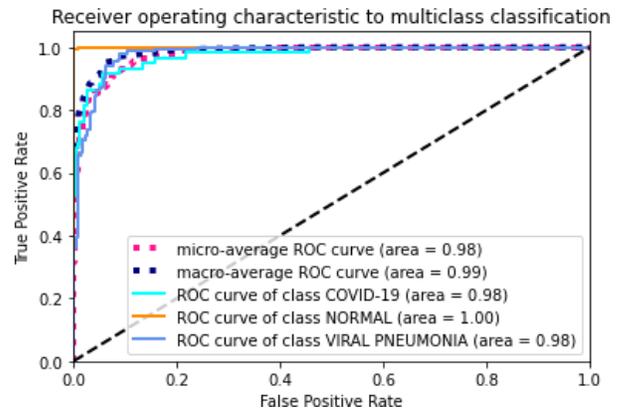

**Fig (a) ROC Curve of VGG - 16 Model**          **Fig (b) ROC Curve of ResNet-50 Model**

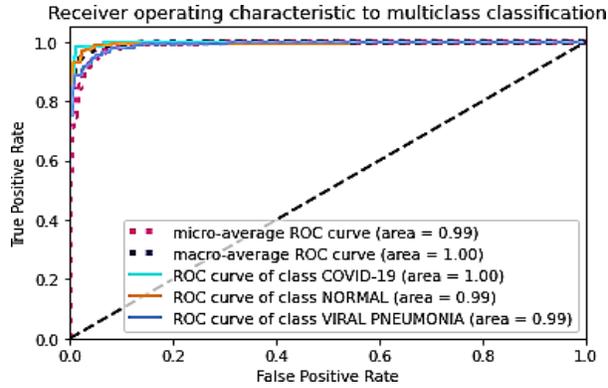 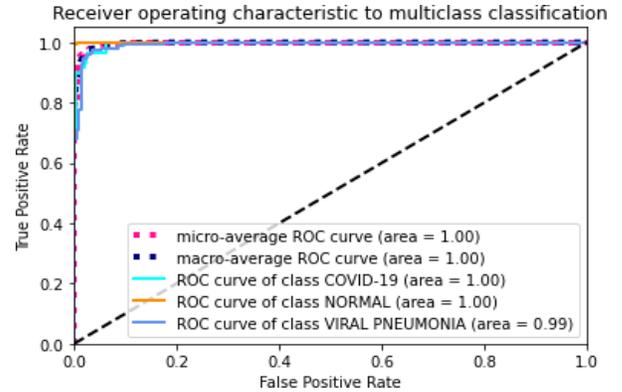

Fig (c) ROC Curve of GoogLeNet Model　　　　　　　　Fig (d) ROC Curve of MNAST Model

Fig [4.7] ROC Curve of Classifiers

AUC - ROC curve is a performance measurement for the classification problems at various threshold settings. ROC is a probability curve and AUC represents the degree or measure of separability. It tells how much the model is capable of distinguishing between classes. Graphically it is plotted with TPR (True Positive Rate) against the FPR (False Positive Rate) where TPR is on the y-axis and FPR is on the x-axis. An excellent model has AUC near to the 1 which means it has a good measure of separability. In our case, almost all four classifiers have AUC equals to 1 or near to 1.

## 4.6 : Discussion

Our research study addressed the issue of overfitting of models which occurs due to small datasets. Like studies [15,16,19,27] we used the same Chest Xray dataset from Cohen's repository. This dataset is publicly available and for our research work we considered the then most updated repository of October 2021. All the images are labelled as COVID – 19, Normal and Viral Pneumonia. Since this research aims to generate synthesized images, so, for image generation, Wasserstein GAN is employed. For classification, four variants of transfer learning techniques are proposed. Among theseVGG-16 model has proved to be the most accurate model which delivered 99.13% accuracy with only 10 epochs. Besides this, MNAST model proved to be the most efficient of all proposed model with 98% accuracy and showing no sign of overfitting.

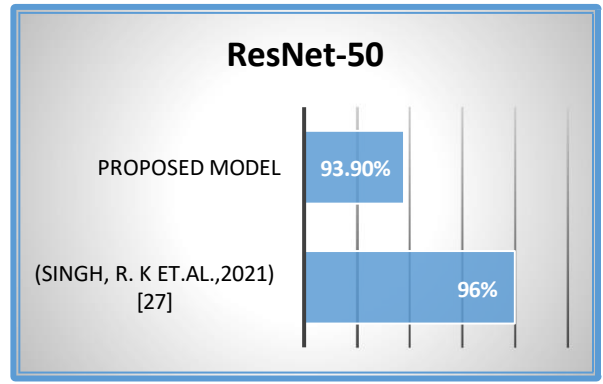

Fig (a) Accuracy Comparison of Proposed ResNet-50 Model with existing model

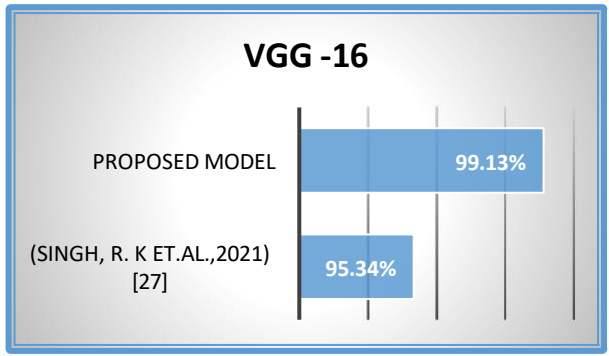

Fig (b) Accuracy Comparison of Proposed VGG -16 Model with existing model

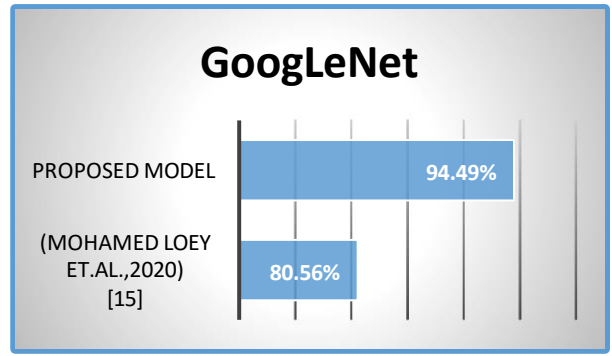

Fig (c) Accuracy Comparison of Proposed GoogLeNet Model with existing model

Fig [4.8] Accuracy comparison of proposed models with the existing models

## 4.7 : Conclusion

Automated medical image diagnosis of COVID-19 cases is a need of time. Numerous research units have already contributed in this context to develop effective methods to prevent and control this pandemic. This research study also aims to contribute effort to develop novel models to detect COVID-19 cases by using deep learning techniques. Main objective of this study is to cater small datasets. Small datasets often lead to model overfitting because deep learning models requires enormous amount of data to train with. To solve this issue, GAN technique used to generate fake and plausible images. In addition to it, this thesis also draws attention towards challenges face by GANs during training. Resolution of those challenges is also considered during image generation. This research study suggests Wasserstein loss to be replaced with BCE loss function to eradicate mode collapse and vanishing gradient. Furthermore, proposed classification models such as VGG-16, ResNet-50, are trained to avoid overfitting. Hyperparameters for these models are taken with care to achieve model generalization.

Following findings can be derived from previous section of results and evaluation:

1. WGAN is a simple architecture which is able to produce plausible images by defeating mode collapse and vanishing gradient issues. Compare with other architectures such as LSGAN[34] and InfoGAN[35] our proposed architecture requires less computation time.
2. Experimental results showed that VGG-16 and Deep CNN model outperformed other models in classifying COVID-19 Chest X-ray images.

Based on the presented study, there are few limitations that lead to following future directions:

1. Evaluating GANs is a particularly challenging task and it is not catered by this presented study. There could be a possibility exist for future researchers to put their effort in evaluating fidelity and diversity of images produce by GANs.
2. To increase diversity with higher resolution images StyleGAN can be used.
3. Design other generative architecture and compare their ability to generate high quality images.